\documentclass[11pt]{article}

\usepackage[a4paper, total={6in, 8in}]{geometry}
\usepackage[english]{babel}
\usepackage[utf8]{inputenc}
\usepackage{amsmath,amssymb}
\usepackage{graphicx}
\usepackage{float}
\usepackage{natbib}
\usepackage{tabularx}
\usepackage{booktabs} 
\usepackage{multirow}
\usepackage{fullpage}
\usepackage{fontawesome}
\usepackage{subcaption} 
\usepackage{amsthm}
\usepackage{comment}
\usepackage{nameref}
\usepackage{appendix} 
\usepackage[dvipsnames]{xcolor}
\usepackage[hidelinks]{hyperref}

\definecolor{darkgreen}{rgb}{0,0.4,0}
\definecolor{darkred}{rgb}{0.5,0.0,0.0}
\definecolor{myred}{HTML}{B22222}   
\definecolor{darkblue}{HTML}{00008B}   
\definecolor{purple4}{RGB}{76,0,153}

\usepackage{pifont} 
\newcommand{\checkvert}{{\color{darkgreen}\ding{51}}}
\newcommand{\nonrouge}{{\color{darkred}\ding{55}}}
\newcommand{\legendmethods}{%
  \textcolor{myred}{red = online, streaming, offline}, 
  \textcolor{blue}{blue = OGK, MCD}, 
  \textcolor{black}{black = sample covariance, oracle}; 
  line types: online = long-dashed, streaming = continuous, 
  offline = dashed, OGK = dot-dash, MCD = continuous, 
  sample covariance = two-dash, oracle = dotted.%
} 

\newcommand{\SR}[2]{\textcolor{gray}{}\textcolor{red}{#2}}

\title{Streaming robust covariance estimation \\
and sequential outlier detection}
\author{Paul Guillot$^{1,2}$, Antoine Godichon-Baggioni$^{1}$, Stéphane Robin$^{1}$ and Laure Sansonnet$^{1,3}$}
\date{}

\begin{document}

\maketitle

\bigskip

\noindent $^{1}$ Sorbonne Université, Université Paris Cité, CNRS, Laboratoire de Probabilités, Statistique et Modélisation, LPSM, F-75005 Paris, France \\ 

\noindent $^{2}$ CREST, CNRS, ENSAE Paris, Institut Polytechnique de Paris, Palaiseau, France \\

\noindent $^{3}$ Université Paris-Saclay, AgroParisTech, INRAE, UMR MIA Paris-Saclay, 91120, Palaiseau,
France
\bigskip

\begin{abstract}
{
Robust estimation of the covariance matrix and outlier detection remain major statistical hurdles, particularly in large datasets where the rate of data contamination increases.
The presence of outliers can severely bias parameter estimates and induce a {\sl masking effect}, whereby some outliers conceal the presence of other outliers, further complicating their detection. Although many approaches have been proposed for covariance estimation and outlier detection, to our knowledge, none of these methods have been implemented in an online setting.
In this paper, we focus on online covariance matrix estimation and outlier detection. Specifically, we propose a new method for simultaneously and online estimating the geometric median and variance, which allows us to calculate the Mahalanobis distance for each incoming data point before deciding whether it should be considered an outlier. 
To mitigate the {\sl masking effect}, robust estimation techniques for the mean and variance are required. Our approach uses the geometric median for robust estimation of the location and the median covariance matrix for robust estimation of the dispersion parameters. The new online methods proposed for parameter estimation and outlier detection allow real-time identification of outliers as data are observed sequentially. The performance of our methods is demonstrated on simulated and real datasets.
}
\end{abstract}

\section{Introduction}

\paragraph{Aim and framework.} 
In the multivariate Gaussian setting, the covariance structure plays a central role in describing data dependence and variability. Classical estimators such as the sample mean and covariance matrix are highly sensitive to outliers, motivating the need for robust alternatives. Indeed, the acquisition of large-scale data in high-dimensional spaces is unfortunately often accompanied by contamination. The automatic detection of these atypical data is not simple, and the use of robust techniques is an interesting alternative, in particular for online outlier detection.
The paper objective is then to adapt and extend robust statistical procedures to the online setting, ensuring both computational efficiency and resistance to contamination.

\paragraph{{Robust estimation of the covariance matrix}.} 

{
Two main approaches can be distinguished for robust covariance estimation.
The first family consists of modifications of the empirical covariance matrix to improve its robustness. This includes, for example, the Minimum Covariance Determinant \citep[MCD:][]{rousseeuw1985multivariate} or shrinkage approaches \citep{Ledoit_Wolf}.
The second family replaces classical covariance and variance estimates with robust alternatives, such as the comedian and the median absolute deviation \citep{falkm}, improvements yielding positive definite estimates \citep[see][]{Cabana_2019,maronna2002robust}.
To the best of our knowledge, none of these methods currently admit an online implementation. The approach we propose is based on the geometric median and the median covariation matrix.
The geometric median is a robust measure of location \citep{haldane1948note,kemperman1987median} that can be preferred to the mean because it has a 50$\%$ breakdown point \citep{gervini2006robust}. It has been extensively studied, and many methods have been proposed to estimate it, both in offline settings \citep{weiszfeld1937point,vardi2000multivariate,beck2015weiszfeld} and online settings \citep{cardot2013efficient,godichon2023online}.
The median covariance matrix, in turn, was introduced by \cite{kraus2012dispersion}, and both offline and online estimation procedures were later proposed by \cite{cardot2017fast}. More recently, \cite{godichon2024robust} used it to reconstruct a robust offline estimator of the variance.
}

\paragraph{Outlier detection.} 
{
Some outlier detection methods rely on dimensionality reduction techniques, assuming that outliers are primarily concentrated along certain principal components: \cite{projfriedmanTukey1974, stahel1981breakdown, donoho1982breakdown, caussinus1990interesting, tyler2009invariant, pena2001multivariate, hubert2005robpca}. 
These approaches are largely distribution-free, but may be computationally demanding, and therefore challenging to implement in an online context.
In contrast, we focus here on Mahalanobis distance–based approaches, which account for the covariance structure of the data \citep{jolliffe1986mathematical}. In this setting, an observation is flagged as an outlier when its associated Mahalanobis distance exceeds a certain threshold. To mitigate the masking effect, robust estimates of location and scatter are required \citep{rousseeuw1990unmasking}. A key advantage of this approach is that it can be efficiently implemented in an online framework.
}

\paragraph{Contribution: A novel online approach.} 
{
In this work, we propose an algorithm to estimate all the quantities of interest (the median, the median covariation matrix, the variance-covariance matrix) in an online manner, while simultaneously enabling online outlier detection.
Although this method is highly accurate, its computational cost can be high if an orthonormalization step is required at each iteration. To address this issue, we introduce a streaming version of our method, which processes data arriving sequentially in batches. This reduces the reliance on costly orthonormalization steps and allows the overall complexity to be reduced to the usual $\mathcal{O}(nd^{2})$ complexity, where $n$ is the number of observations and $d$ is the data dimension. Our method also enables real-time detection of outliers in incoming data without the need to recompute previous estimates from scratch.
}

\paragraph{Outline.} 
{
The paper is organized as follows.
Section~\ref{sec:offline} reviews existing offline procedures, including the geometric median and the median covariation matrix.
The online and the streaming versions of the novel algorithm are described in Section~\ref{sec:online}. Finally, Section~\ref{sec:simu} presents numerical experiments that assess the performance and accuracy of the proposed method, alongside an application to a real-world dataset. The proposed online and streaming methods are implemented in the \texttt{STARRS} package, which is publicly available on CRAN (\url{https://cran.r-project.org/web/packages/STARRS/index.html}).
}

\section{Classical offline methods}
\label{sec:offline}

In this section, we provide more details about
the principal robust approaches for covariance matrix estimation and outlier detection, with particular attention to their existing formulations in an online framework.

\subsection{Robust estimation of the covariance matrix}
\label{subsec:estcov}

\paragraph{Sample covariance based methods.}\label{par:covariance}
The sample mean and covariance matrix are highly sensitive to outliers, motivating robust alternatives based on modified covariance estimators. The minimum covariance determinant (MCD) estimator \citep{rousseeuw1985multivariate}, in the context of elliptical distributions, seeks the subset of $h$ observations with the smallest covariance determinant, yielding robust estimates of multivariate location and scatter. However, its exact computation is combinatorial, and the FAST-MCD algorithm \citep{rousseeuw1999fast} was proposed to approximate it efficiently via iterative C-steps. As far as we know, no online version currently exists. Other approaches, in the context of functional data, include the trimmed covariance estimator \citep{gervini2012outlier}, which discards identified outliers, and the shrinkage estimator \citep{Ledoit_Wolf}, which combines the sample covariance with a structured target to improve robustness.
FAST-MCD runs in $\mathcal{O}(nd^2 + d^3)$ per concentration step \citep{rousseeuw1999fast}, and typically requires several hundred such steps, so its overall cost remains substantially higher than that of a one-pass procedure. Since the optimal $h$-subset depends globally on the data, an online implementation would require 
recomputing it from scratch at each new observation, which is impractical.

\paragraph{Comedian-based methods.}\label{par:comedian}
Another family of methods for estimating location and scatter parameters relies on the \textit{Comedian} approach, which replaces classical covariance and variance with robust alternatives: the \textit{Comedian} ($\operatorname{COM}$) and the \textit{squared Median Absolute Deviation} ($\operatorname{MAD}^2$). For random variables $U$ and $V$, these are defined as  
$
\operatorname{COM}(U, V) = \operatorname{med}\big[(U - \operatorname{med}(U))(V - \operatorname{med}(V))\big]
$, and
$
\operatorname{MAD}^2(U) = \operatorname{med}\big[(U - \operatorname{med}(U))^2\big],
$
where $\operatorname{med}$ denotes the median \citep{falkm}.  
The corresponding \textit{Comedian} matrix $S_C$ for a dataset $\mathbf{X} \in \mathcal{M}_{n,d}(\mathbb{R})$ is defined by 
$
S_C(i,j) = \operatorname{COM}(\mathbf{X}[,i],\, \mathbf{X}[,j]).
$
Although $S_C$ provides a robust covariance estimate, it is not necessarily positive definite. To address this limitation, \citet{Cabana_2019} proposed a shrinkage correction inspired by \citet{Ledoit_Wolf}, producing a robust and positive definite covariance estimator.
The Orthogonalized Gnanadesikan–Kettenring (OGK) estimator \citep{maronna2002robust} offers a related strategy: it uses the robust pairwise covariance identity of \citet{gnanadesikan1972robust}, substitutes classical variances with robust scale estimators, and applies an orthogonalization step that standardizes the data, projects it onto the eigenvectors of the intermediate scatter matrix, and re-estimates variances along these principal directions. Despite its strong robustness and computational efficiency, an online implementation remains unavailable. The offline algorithm operates with a complexity of $O(n d^2 + d^3)$ \citep{maronna2002robust}. 
\paragraph{Offline median covariation matrix based method.} \label{par:varreconstruct}  
Our covariance matrix estimation is founded on the estimation of the geometric median and the median covariation matrix (MCM) described in Appendix~\ref{appendix:geommed_mcm}. In the case where the distribution of $X$ is symmetric, the MCM and the usual variance share the same eigenvalues \citep{kraus2012dispersion}, and one has to give a link between their eigenvalues to be able to reconstruct the variance from the MCM (in a robust way), which is the purpose of Section~\ref{subsec:reconstructoffline}.

\subsection{Outlier detection}\label{subsec:outldet}

\SR{We highlight here two main families of outlier-detection methods. 
Dimension-reduction–based approaches (paragraph~\nameref{par:dimension_red_based}) rely on projecting the data onto informative low-dimensional subspaces, possibly using robust estimates of location and scatter and robust Mahalanobis distance, to reveal anomalous observations. In contrast, distance-based methods (paragraph~\nameref{par:maha}) focus on obtaining a robust estimate of the covariance matrix and subsequently detect outliers using robustified Mahalanobis distances, as their sole detection tool.}{}

\paragraph{Dimensionality reduction based methods.} \label{par:dimension_red_based} 
Dimension-reduction–based outlier detection methods aim to identify projection directions along which outliers become most distinguishable. Although some of these approaches use or produce robust estimates of location or scatter, robust variance estimation is typically not their primary objective, and outliers are often assumed to be sparse. After projecting the data, a suitable metric is applied to flag anomalous points. In contrast to the MCD and to our proposed methods, these approaches generally do not rely on explicit distributional assumptions on the data. Several methodological variants have been developed within this projection–based framework. \\
The most classical dimension-reduction technique is Principal Component Analysis (PCA), which identifies directions of maximal variance \citep{pearson1901liii,hotelling1933analysis,Jolliffe2002}. 
However, standard PCA neither uses robust scale estimates nor protects against the masking effect, and its initial purpose is not outlier detection. \cite{hubert2005robpca} later introduced ROBPCA, a robust PCA method suitable for high-dimensional data and based on the MCD estimator. \cite{filzmoser2008outlier} developed another principal component-based robust procedure effective in high-dimensional settings. \\
A key limitation of most of the PCA-based approaches is that they mainly detect observations that inflate the variance. \cite{projfriedmanTukey1974} introduced the notion of \emph{projection pursuit}, which seeks projections that optimize a projection index designed to reveal non-Gaussian structure. Their original index is not robust to outliers, and evaluating many projections is computationally demanding. Robust projection indices were later proposed by \cite{stahel1981breakdown,donoho1982breakdown}, based on the mean absolute deviation and computed for each observation. Peña and Prieto \cite{pena2001multivariate} further suggested using directions that maximize or minimize kurtosis. Although more robust, these methods remain computationally intensive. Finally, the invariant coordinate selection (ICS) method \citep{tyler2009invariant,caussinus1990interesting}, based on the joint diagonalization of two scatter matrices, is effective for detecting a small number of outliers. Overall, despite their robustness properties, most projection-based methods require exploring many directions and thus remain challenging to apply in real-time or streaming settings.

\paragraph{Mahalanobis distance based outlier method.} \label{par:maha}
In a multivariate setting, graphical methods for outlier detection are often inefficient. The squared Mahalanobis distance 
$$
D_i = (X_i - \mu)^{\top} \Sigma^{-1}(X_i - \mu)
$$
offers a more suitable alternative by taking into account the covariance structure of the data,  which is an advantage compared to the classical Euclidean distance. After estimating the location parameter $\mu$ and the scatter parameter $\Sigma$, the corresponding estimators $\widehat{\mu}$ and $\widehat{\Sigma}$ are used to compute an estimate of the squared Mahalanobis distance $D_i$ for each observation.  Then, an observation $X_i$ is classified as an outlier if $\widehat{D}_i = (X_i - \widehat{\mu})^\top \, \widehat{\Sigma}^{-1} \, (X_i - \widehat{\mu}) > t$, where $t$ denotes a given threshold.
This criterion’s primary advantage is its straightforward online implementation (see Section~\ref{subsec:onloutldet}). Our procedure avoids the explicit computation of the inverse of $\widehat{\Sigma}$, that can be computationally expensive.
Denoting $\widehat{m}$ an estimate of the geometric median $m$, $\widehat{\lambda}$ an estimate of the vector $\lambda$ containing the eigenvalues of $\Sigma$ and $\widehat{P}$ the matrix containing the associated eigenvectors of an estimate of the eigenvectors of $\Sigma$, we avoid direct computation of $\Sigma^{-1}$ and $\widehat{D}_i$ is computed as
$$
\widehat{D}_{i} = \sum_{j=1}^{d} \frac{1}{\widehat{\lambda}[j]}  \big\langle X_{i} - \widehat{m}, \widehat{P}  [,j] \big\rangle^{2}.
$$
When $\Sigma$ and $\mu$ are known, the Mahalanobis distance $D_i$ follows a $\chi^2(d)$-distribution. When $\Sigma$ and $\mu$ are estimated by the sample mean $\overline{X}_{n}$ and the sample covariance matrix $S$, the estimated distance $\widehat{D}_i$ follows a scaled $F$-distribution: $\widehat{D}_i \sim \frac{n+1}{n} \cdot \frac{d(n - 1)}{n - d} F(d, n - d)$, where $F(d, n - d)$ denotes the Fisher distribution with $d$ and $n - d$ degrees of freedom \cite[see][]{hotelling1931generalization}. When the sample size $n$ is large, the scaling factor $\frac{n+1}{n} \cdot \frac{d(n - 1)}{n - d}$ becomes close to $d$, and the distribution of $\widehat{D}_i$ approaches $\chi^2(d)$. However, {replacing the empirical covariance matrix with a robust version can cause deviations from the chi-squared distribution}. Several methods have been proposed to address this issue. A common approach is to scale the distances by a constant factor. For example, \cite{maronna2002robust} proposed to scale the squared Mahalanobis distances by the factor $\frac{\chi^2_d(0.5)}{\operatorname{med}(\widehat{D}_1, \ldots, \widehat{D}_n)}$  
and define the scaled Mahalanobis distance
\begin{equation} \label{eq:scaledMaha}
\widetilde{D}_i = \frac{\chi^2_d(0.5)}{\operatorname{med}(\widehat{D}_1, , \ldots, \widehat{D}_n)} \widehat{D}_i.
\end{equation}
This scaling makes the empirical median of the corrected distances match with the median of the $\chi^2(d)$-distribution.
Note that this median-based correction is used in our online method 
(see Section~\ref{sec:online}), whereas OGK and MCD rely on the nominal 
threshold $t = \chi^2_d(0.95)$. The correction assumes that the empirical distribution of 
Mahalanobis distances is a scaled version of the $\chi^2_d$ law. This assumption 
is violated by OGK, whose distances are compressed in the body but as heavy as 
$\chi^2_d$ in the tail, so that no single scaling factor can align both the 
median and the $95\%$-quantile; it also fails for MCD, which satisfies the shape 
assumption but relies on a median contaminated by the outliers.

\subsection{Details on the MCM based offline reconstruction of the variance} \label{subsec:reconstructoffline}

We now introduce the offline estimation of the median covariation matrix. To this aim,
for any vector $Z \in \mathbb{R}^{d}$, we denote by $Z[i]$ its $i$-th component for any $i=1, \ldots,d$. In addition, let us recall that $X \sim \mathcal{N}(\mu, \Sigma)$.  Let us denote by $ \lambda = (\lambda[k])_{k = 1 , \ldots ,d} $ the vector of eigenvalues of $ \Sigma $, and by $ \delta = (\delta[k])_{k =1, \ldots,d}$ the vector of eigenvalues of the median covariance matrix. The relationship between $ \delta $ and $ \lambda $ (see Proposition 2 in  \cite{kraus2012dispersion}) is given by 
\begin{equation}\label{relationdeltalambda}
\delta =  \dfrac{\mathbb{E} \left[ \frac{\lambda \odot U^{\otimes 2}}{h\left(\lambda,\delta,U \right)} \right] }{ \mathbb{E} \left[ \frac{1}{h\left(\lambda,\delta,U \right)} \right]},
\end{equation}
where $U \sim \mathcal{N}(0,I_{d})$, $ \lambda \odot U $ denotes the Hadamard product of the vectors $ \lambda $ and $ U $, $U^{\otimes 2} = U \odot U$ and
\[
h(\lambda, \delta, U) := \sum_k (\delta[k] - \lambda[k] U[k]^2)^2 + \sum_{i \neq j} U[i]^2 U[j]^2  \lambda[i] {\lambda[j]}.
\]
The relation between $\delta$ and $\lambda$ given by \eqref{relationdeltalambda} allows us to reformulate the problem of finding $ \lambda $ as the search of the zero of a function. More precisely, denoting $ A = \operatorname{diag}(U[1]^2, \ldots, U[d]^2) $, the objective is to determine $ \lambda $ such that 
\begin{equation}\label{relRM}
\mathbb{E}\left[ \frac{A \lambda - \delta}{\sqrt{h(\lambda, \delta, U)}} \right] = 0.    
\end{equation}
Then, the aim is to estimate $\lambda$ with the help of a Robbins-Monro procedure coupled with Monte-Carlo method. More precisely, let us generate  i.i.d.\,copies $ U_1 , \ldots, U_{n}, \ldots$  of $ U  $ and at each new time $n+1$, denote $ A_{n+1} = \operatorname{diag}(U_{n+1}[1]^2, U_{n+1}[2]^2, \ldots, U_{n+1}[d]^2) $. Then, the estimates of $\lambda$ are defined recursively for all $n \geq 0$ by   
\begin{equation}\label{def::lambda}
\lambda_{n + 1} = \lambda_n - \gamma_{n + 1} \frac{A_{n+1} \lambda_n - \delta}{\sqrt{h(\lambda_n,\delta,U_{n+1})}}
\end{equation}
where $\lambda_0$ is chosen arbitrarily, and $\gamma_{n}=c_{\gamma}(n+n_{0})^{-\gamma}$ with $c_{\gamma} > 0$, $\gamma\in (1/2,1)$ and $n_{0} \geq 0$ \citep{robbins1951stochastic,godichon2024robust}.
Unfortunately, the Robbins-Monro algorithm cannot achieve asymptotic efficiency, so a common method is to consider its averaged (or weighted) version recursively defined for all $n \geq 0$  by  
\begin{equation}\label{def::lambdabar}
\overline{\lambda}_{n+1} = \overline{\lambda}_n + \frac{\operatorname{log}(n+1)^{\omega}}{\sum_{\ell = 0}^n \operatorname{log}(\ell+1)^{\omega}} (\lambda_{n+1} - \overline{\lambda}_n)  
\end{equation}
where $\overline{\lambda}_{0} = \lambda_{0}$ and $\omega \geq 0$  \citep{mokkadem2011generalization, boyer2023asymptotic, godichon2024robust}.
Then, denoting by $n_{MC}$ the total number of generated data $U_{n}$ (and iterations so), the estimate $\overline{\lambda}_{n_{MC}}$ is totally defined by   $\delta, \lambda_{0}, n_{MC}$ and $n_{0}$. The MCM also provides access to the eigenvectors of the covariance matrix. Denoting by $P$ the matrix whose columns are the eigenvectors obtained from the MCM, the covariance matrix can be estimated as $ P \Delta P^{\top}$, where $\Delta$ is a diagonal matrix containing the estimated eigenvalues $\overline{\lambda}_{n_{\text{MC}}}$ of $\Sigma$. We give an online implementation of this method in Section~\ref{sec:online}.

\subsection{Summary} \label{subsec:summary}

Table~\ref{tab:online_cov_methods} summarizes the methods discussed above, indicating whether their primary objective is the robust estimation of the covariance matrix and whether outlier identification relies on dimension reduction or exclusively on a (robust) Mahalanobis-distance–based criterion.

\begin{table}[h!]
\centering
\caption{Summary of existing methods. \label{tab:online_cov_methods}}
\renewcommand{\arraystretch}{1.5}
\resizebox{\textwidth}{!}{
\begin{tabular}{@{}lcccc@{}}
\toprule
\textbf{Method} & 
\shortstack{\textbf{Dimension} \\ \textbf{reduction}} & 
\shortstack{\textbf{Robust Mahalanobis} \\ \textbf{distance}} & 
\shortstack{\textbf{Online?}} \\ 
\midrule
PCA \citep{pearson1901liii,hotelling1933analysis, Jolliffe2002}  & \checkvert & \nonrouge & \nonrouge \\
Robust PCA \cite{hubert2005robpca}  & \checkvert & \nonrouge & \nonrouge \\
PP \citep{projfriedmanTukey1974}  & \checkvert & \nonrouge & \nonrouge \\
PP \citep{stahel1981breakdown, donoho1982breakdown}  & \checkvert & \nonrouge & \nonrouge \\
PP kurtosis \citep{pena2001multivariate}   & \checkvert & \nonrouge & \nonrouge \\
Invariance coordinate selection \citep{caussinus1990interesting, tyler2009invariant} & \checkvert & \nonrouge & \nonrouge \\
Trimmed covariance estimator \citep{gervini2012outlier}   & \nonrouge & \checkvert & \nonrouge \\
Shrinkage \citep{Ledoit_Wolf}  & \nonrouge & \checkvert & \nonrouge \\
Shrinkage \citep{Cabana_2019}  & \nonrouge & \checkvert & \nonrouge \\
OGK \citep{maronna2002robust}  & \nonrouge & \checkvert & \nonrouge \\
MCD \citep{rousseeuw1985multivariate}  & \nonrouge & \checkvert & \nonrouge \\
MCM-based \citep{godichon2024robust} & \nonrouge & \checkvert & \nonrouge \\
\textbf{This work} & \nonrouge & \checkvert & \checkvert \\
\bottomrule
\end{tabular}
}
\end{table}

\section{Our novel online approach}
\label{sec:online}

In the following, we first present the online covariance estimation and outlier detection procedure based on the sample covariance estimator. This approach will serve as a benchmark for our methods as, to the best of our knowledge, no other online procedures capable of performing both tasks simultaneously currently exist. We then introduce our proposed methodology in both the online and batch settings.

\subsection{Sample mean and sample covariance online method} \label{subsec:onlinesamplecov}

A first naive online approach would be to consider online estimates of the mean and the variance based on the sample mean and the sample covariance matrix. More precisely, after initializing the mean and covariance estimators using the sample mean and covariance computed from the first (for instance, 100) observations, given a new data $X_{n+1}$, the estimates can be updated as follows: 
\begin{align*}
 & \overline{X}_{n+1}  = \overline{X}_n + \frac{1}{n+1} (X_{n+1} - \overline{X}_n),  \\
 & 
\Sigma_{n+1} =  \Sigma_n + \frac{1}{n+1} \left( X_{n+1} - \overline{X}_{n} \right) \left( X_{n+1} - \overline{X}_{n} \right)^\top   
\end{align*}
Subsequently, we estimate the inverse of $\Sigma_{n + 1}$ with the following update based on the Sherman-Morrisson formula: 
\[
  \Sigma_{n+1}^{-1} =  \frac{n+1}{n}  \Sigma_{n}^{-1} - \frac{n+1}{n^{2} }\frac{1}{1 + U_n^{\top}   n^{-1}\Sigma_n^{-1}  {U}_n} \Sigma_{n}^{-1}  {U}_n   {U}_n^{\top}  \Sigma_{n}^{-1}  
\]
where $U_n = X_n - \overline{X}_{n-1}$. 
 One can then calculate the Mahalanobis distance of the new data as
\[
\widehat{D}_{n+1} = (X_{n+1} - \overline{X}_{n})^\top   \Sigma_{n+1}^{-1} (X_{n+1} - \overline{X}_{n+1}) .
\]
Finally, a new observation is flagged as an outlier if $\widehat{D}_{n+1} > t$, where $t$ is a predefined threshold. As in the offline setting, this method lacks robustness to outliers, which can significantly distort both the covariance estimates and the outlier detection process.

\subsection{Median covariation matrix based method in an online setting} \label{subsec:mcmonline}

In the sequel, we consider i.i.d.\,copies $X_{1} , \ldots ,X_{n}, X_{n+1}, \ldots $ arriving sequentially. We then propose a new method, based on the ideas proposed in Section~\ref{subsec:estcov}, to both estimate online the geometric median, the MCM and a robust estimate of the variance. This also allows to calculate at each step the new Mahalanobis distance. More precisely, when a new data $X_{n+1}$ arrives, one can make the following scheme:
\begin{align*}
m_{n+1} &= m_n + \gamma_{n+1} \frac{X_{n+1} - m_n}{||X_{n+1} - m_n||}  \\
\overline{m}_{n+1} &= \overline{m}_n +  \frac{1}{n + 2}(m_{n+1} - \overline{m}_n ) \\
V_{n+1} &= V_n + \gamma_{n+1} \frac{(X_{n+1} - \overline{m}_n)(X_{n+1} - \overline{m}_n)^\top - V_n}{||(X_{n+1} - \overline{m}_n)(X_{n+1} - \overline{m}_n)^\top - V_n||_F} \\
\overline{V}_{n+1} &= \overline{V}_n +  \frac{1}{n + 2} \left( V_{n+1}- \overline{V}_n \right) \\
\SR{\text{P}_{n+1} &= \operatorname{eigen}(\overline{V}_{n+1}) \\}{} 
\SR{\delta_{n+1} &= \operatorname{Sp}(\overline{V}_{n+1}) \\}{} 
\operatorname{\lambda_{n+1}} &= \operatorname{RM}\left( \delta_{n+1}, \lambda_n, n_{MC}, n \times n_{MC}    \right)  \\
\widehat{D}_{n+1} &= \sum_{j=1}^{d} \frac{1}{\lambda_{n+1}[j]} \big\langle X_{n+1} - \overline{m}_{n+1} , \text{P}_{n+1}[,j]  \big\rangle^{2}  \\
\end{align*}
{where $P_{n+1}$ is the matrix formed by the eigenvectors of $\overline{V}_{n+1}$ and $\delta_{n+1}$ is the set of its eigenvalues.}

The estimates $m_{n}$ and $\overline{m}_{n}$ (resp.\,$V_{n}$ and $\overline{V}_{n}$) correspond to the stochastic gradient algorithm, with $\gamma_{n} = c_{\gamma} (n + n_0)^{-\gamma}$ with $c_{\gamma} > 0$ and $n_0 \geq 0$, and its averaged version 
\citep{cardot2013efficient} for estimating the geometric median (resp.\,the MCM, 
see \cite{cardot2017fast}).
Then, it allows to obtain (an approximation of) the eigenvalues  (resp\,eigenvectors) of the estimate of the MCM, denoted by  $\delta_{n+1}$ (resp. $\text{P}_{n+1}$). Next, we denote $\operatorname{RM}(.,.,.,.)$ the random function linking $\delta,\lambda_{0},n_{MC}$ and $n_{0}$ to $\lambda_{n_{MC}}$ and $\overline{\lambda}_{n_{MC}}$ according to Equations \eqref{def::lambda} and \eqref{def::lambdabar}.   
Finally, one can eventually update a robust estimate of the covariance matrix given by 
$$
\hat{\Sigma}_{n+1} = \text{P}_{n+1} \text{diag} \left( \lambda_{n+1} \right) \text{P}_{n+1}^{\top}.
$$

\paragraph{Initialization.} 
In a practical way, in order to initialize the different estimates, we use the offline method on the $n_{\text{init}}$ first data (with $n_{\text{init}}$ chosen arbitrarily equal to $500$ in the simulations) to obtain estimates $m_{n_{\text{init}}}$ (resp.\,$ V_{n_{\text{init}}}$) before taking $\overline{m}_{n_{\text{init}}} = m_{n_{\text{init}}}$ (resp.\,$\overline{V}_{n_{\text{init}}} = V_{n_{\text{init}}}$). We then obtain first estimates $\delta_{n_{\text{init}}}$ (resp.\,$\operatorname{EV}_{n_{\text{init}}}$) of the eigenvalues (resp.\,eigenvectors) of the MCM. We then apply the Robbins-Monro algorithm to obtain $\lambda_{n_{\text{init}}} = \operatorname{RM}(\delta_{n_{\text{init}}},\lambda_{0},n_{\text{init}} \times n_{MC} , 0)$ with $\lambda_{0} = \delta_{n_{\text{init}}}$  for obtaining a first estimate of the eigenvalues of the variance.  

\paragraph{Parameters.} 
Concerning the hyperparameters, we choose to take $\gamma_{n}$ of the form $\gamma_{n} = c_{\gamma}n^{-\gamma}$ with $c_{\gamma}> 0$ and $\gamma \in (1/2,1)$. Observe that one could take different $\gamma_{n}$ to update $m_{n}$ and $V_{n}$, see 
\cite{cardot2017fast}. In addition, $n_{MC}$ corresponds to the number of data generated at each time for the Robbins-Monro procedure.

\paragraph{Computational complexity.} 
Updating the median estimates requires $\mathcal{O}(d)$ operations per update, compared to $\mathcal{O}(d^2)$ for the MCM. Furthermore, obtaining the eigenvalues and eigenvectors, as well as potentially updating the variance, unfortunately scales to $\mathcal{O}(d^{3})$ operations.
Then, the reconstruction of the eigenvalues of the variance necessitates $\mathcal{O}(n_{MC}d^{2})$ operations. Finally, the computation of the Mahalanobis distance has a complexity of order $\mathcal{O}(d^{2})$.
Specifically, for $N$ data points, the overall computational complexity is
$$
\underbrace{\mathcal{O}(N d)}_{\text{Updating } m_{n} \text{ and } \overline{m}_{n} } +
\underbrace{\mathcal{O}(N d^2)}_{\text{Updating } V_{n} \text{ and } \overline{V}_{n} } +
\underbrace{\mathcal{O}(N d^3)}_{\text{Eigen decomposition}} +
\underbrace{\mathcal{O}(N n_{\text{MC}}d^{2})}_{\text{Updating } \lambda_{n}} + 
\underbrace{\mathcal{O}(N d^2)}_{\text{Calculate } \widehat{D}_{n+1}}
$$
Then, as soon as one can chose $n_{MC}$ arbitrarily, the main cost comes from the spectral decomposition at each step. The aim is so to reduce the frequency of the use of this spectral decomposition, leading to the streaming (also called online mini-batch) framework.

\subsection{Median covariation matrix based method in a batch setting} \label{subsec:mcbatch}

In this framework, we consider data arriving by block of size $s$, or one can do it artificially. More precisely, at time $n+1$, we consider new i.i.d.\,copies 
 $\{X_{n+1,j}\}_{j=1,\ldots,s}$ treated as a single statistical unit. The main change is that we now consider streaming (or online mini-batch) algorithms for estimating the median and the MCM, i.e.\,that we consider averaged estimates (based on the new block of data) of the gradients, leading to the following updates \citep{Godichon_Baggioni_2023}:
\begin{align*}
 m_{n + 1} &= m_n + \gamma_{n+1} \frac{1}{s} \sum_{j = 1}^s \frac{X_{n+1 ,j} - m_n}{\|X_{n+1,j} - m_n\|} \\
 \overline{m}_{n+1} &= \overline{m}_n + \frac{1}{n + 2}(  m_{n+1} - \overline{m}_n) \\
 V_{n + 1} &= V_n + \gamma_{n+1} \frac{1}{s} \sum_{j = 1}^s \frac{(X_{n+1,j} - \overline{m}_n)(X_{n+1 ,j} - \overline{m}_n)^\top - V_n}{\|(X_{n+1, j} - \overline{m}_n)(X_{n+1,j} - \overline{m}_n)^\top - V_n\|_F} \\
 \overline{V}_{n+1} &= \overline{V}_n + \frac{1}{n + 2} \left(  V_{n+1} - \overline{V}_n \right).
 \end{align*}
Observe that in this case, we take $\gamma_{n} = \sqrt{s}c_{\gamma}n^{-\gamma}$ to take into account the fact that we do less iterations. The updates of $\delta_{n+1},\text{EV}_{n+1},\hat{\Sigma}_{n+1}$ are the same as for the online setting, while there is a little modification for $\lambda_{n+1}$ consisting in taking $\operatorname{\lambda_{n+1}} = \operatorname{RM}\left( \delta_{n+1}, \lambda_n, n_{MC}, n \times s \times  n_{MC}\right) $. Finally, one can then calculate the Mahalanobis distance of the new data, i.e.\,for all $i=1 , \ldots , s$,
\begin{align*}
\widehat{D}_{n+1,i} &= \sum_{j=1}^{d} \frac{1}{\lambda_{n+1}[j]} \big\langle X_{n+1,i} - \overline{m}_{n+1} , \text{P}_{n+1}[,j]  \big\rangle^{2}    .
\end{align*}
 The main advantage of this method is that if we denote by $N$ the total number of data dealt with, we only do  $N/s$ iterations. This means that we reduce the number of time we use the costly spectral decomposition, leading to a total calculus complexity of order
 $$
\underbrace{\mathcal{O}(N d)}_{\text{Updating } m_{n} \text{ and } \overline{m}_{n} } +
\underbrace{\mathcal{O}(N d^2)}_{\text{Updating } V_{n} \text{ and } \overline{V}_{n} } +
\underbrace{\mathcal{O}\left( N d^3 / s  \right)}_{\text{Eigen decomposition}} +
\underbrace{\mathcal{O}(N n_{\text{MC}}d^{2})}_{\text{Updating } \lambda_{n}} + 
\underbrace{\mathcal{O}(N d^2)}_{\text{Calculate all } \hat{D}_{n+1,i}}
$$
Then, taking $s=d$ can lead to a total complexity of order $\mathcal{O}(N n_{\text{MC}}d^{2})$, which is (up to $n_{MC}$) the same calculus time as the naive method. Observe that one can initialize in the same way  as in  the online case.

\subsection{Online outlier detection} \label{subsec:onloutldet}

\paragraph{Specificity and advantages of our method.}
Our proposed method performs, simultaneously, robust covariance estimation in the presence of outliers and online outlier detection. This brings two key advantages. First, observations flagged as outliers can be identified and handled immediately upon arrival. Second, the procedure is fully online and does not require storing past data, and recomputation from scratch. 

\paragraph{Outlier detection procedure.} 
Subsequently, once the Mahalanobis distance $\widehat{D}_{n+1}$ of the new observation $X_{n+1}$ has been computed, the observation is flagged as an outlier whenever the scaled Mahalanobis defined in Equation~\eqref{eq:scaledMaha} exceeds  a predefined threshold $t$. The scaling factor
\SR{$\chi^2_d(0.5) \big/ \operatorname{med}(\widehat{D}_1,\ldots,\widehat{D}_n)$}{}
requires online estimation of the median of past Mahalanobis distances {$\operatorname{med}(\widehat{D}_1,\ldots,\widehat{D}_n)$}. We update this quantity using the classical Robbins–Monro stochastic approximation scheme (see \citealp{robbins1951stochastic, labopin2016methodes}). Denoting by $\text{med}_n$ the estimate of $\operatorname{med}(\widehat{D}_1,\ldots,\widehat{D}_n)$ at iteration $n$, the update rule is
\[
\text{med}_{n+1} 
= \text{med}_n 
- \gamma_{n+1}\big(\mathbf{1}_{\{\widehat{D}_{n+1} \le \text{med}_n\}} - 0.5\big),
\]
where $\gamma_n = c_{\gamma}(n + n_0)^{-\gamma}$ .

\section{Simulation}\label{sec:simu}

\paragraph{Aim.} 
We evaluate our proposed algorithms with two primary goals: accurately estimating the true covariance matrix  $\Sigma$  despite outliers, and efficiently detecting those outliers.

\paragraph{Algorithms.}
In what follows, we consider both online and offline estimation procedures. 
The online procedures are:
(i)~the \textit{sample covariance online method}, 
which refers to the online estimation of the sample covariance matrix, see 
Section~\ref{subsec:onlinesamplecov};
(ii)~the \textit{online method}, which stands for the median covariation matrix-based approach in pure online  processing (batches of size~1, Section~\ref{subsec:mcmonline}); and 
(iii)~the \textit{streaming method}, which designates the median covariation matrix approach in batch processing (with batches of size $s=10$, 
 Section~\ref{subsec:mcbatch}).
In addition, the offline 
procedures consist of our proposed (iv)~\textit{offline} method based on the median covariation matrix (MCM), together with the two competing offline methods 
(v)~\textit{OGK} and (vi)~\textit{MCD} (see Section~\ref{sec:offline}).  As a theoretical benchmark, 
we also consider an (vii) \textit{Oracle} estimator, defined as the Mahalanobis 
distance computed with the true mean and covariance matrix of the 
uncontaminated population. 
\paragraph{Implementation.}

Our proposed methods, the \textit{offline} MCM based method, and the \textit{online}  and  \textit{streaming}  methods  are  implemented  in  the \texttt{STARRS}  package,  which  is  publicly  available in CRAN (\url{https://cran.r-project.org/web/packages/STARRS/index.html}).
 For \textit{OGK} and \textit{FAST-MCD} covariance estimators, we used the publicly available \texttt{robustbase} package. Note that the \textit{online sample covariance method} was implemented in \texttt{R} and \texttt{Rcpp}~1.0.9.

\subsection{Simulation design} \label{subsec:simdesign}

\paragraph{Distributions.}  
To mimic the distribution of contaminated data, we used the following mixture model:  
\[
(1 - r) \mathcal{F}_0 + r \mathcal{F}_1,
\]
where $\mathcal{F}_0$ stands for the reference distribution, $\mathcal{F}_1$ for the outlier distribution, and $r$ for the contamination rate. 
For the reference distribution, we used a multivariate Gaussian distribution in $\mathbb{R}^d$: $\mathcal{F}_0 = \mathcal{N}\left( \mu_0, \Sigma_0\right)$, with $\mu_0 = \mathbf{0}_d$ and  $\Sigma_0 = D_0 T_0 D_0$. We considered heterogeneous variances, taking $D_0 = \text{diag}(\sigma_{0, 1}, \ldots, \sigma_{0, d})$ with $\sigma^2_{0,i} = 2 i / (d+1)$ and correlated coordinates, taking $T_0$ as a Toeplitz matrix with entries $(T_0)_{ij} = \rho_0^{|i-j|}$, with $\rho_0 = 0.3$. 
We also considered a multivariate Gaussian distribution for the contamination distribution: $\mathcal{F}_1 = \mathcal{N}\left(\mu_1, \Sigma_1\right)$ parametrized as follows
\begin{itemize}
    \item $\mu_1 = \mu_0 + k m_1$ with $k \geq 0$ and $m_1 = ((-1)^1, \dots, (-1)^d)^{\top}$,
    \item $\Sigma_1 = D_1 T_1 D_1$, where
    \begin{itemize}
        \item[$\circ$] $D_1 = \ell D_0$, with $\ell > 0$, and
        \item[$\circ$] $T_1$ is a Toeplitz matrix with entries $(T_1)_{ij} = \rho_1^{|i-j|}$, with $\rho_1 \in (-1, 1)$.
    \end{itemize}
\end{itemize}
The three tuning parameters $k$, $\ell$ and $\rho_1$ control the mean shift, the variance scaling and the correlation structure (covariance orientation), respectively.

\paragraph{Simulation scenarios.} 
The simulation framework evaluates covariance matrix estimation and 
anomaly detection methods under four geometric transformations of the data 
distribution: dilatation, location shift, concentration, and deformation 
(see Table~\ref{tab:contscentransf}). The first two alter the scale and 
location of the contaminating distribution, whereas the last two modify its 
shape. Because concentration and deformation exhibit similar qualitative behaviors, 
they are treated jointly; however, to maintain conciseness, the results for concentration are presented in this section, while those for deformation which are deferred to  Appendix~\ref{app:deformation}.
\begin{description}
\item[{\sl Dilatation.}]
Dilatation corresponds to an expansion of the data dispersion, induced by $\ell > 1$. This increases variability and leads to anomalies that are more spread in the feature space.
\item[{\sl Location shift.}]
Controlled by the parameter $k$, the location shift determines the displacement of the anomaly center, thereby enabling both centered and shifted contamination regimes.
\item[{\sl Concentration and deformation.}]
Both transformations act on the shape of the covariance structure rather than 
on its location or magnitude. Concentration is governed by a scaling parameter 
$\ell < 1$, which compresses the scatter matrix around the central tendency and 
produces highly concentrated anomaly clusters. Deformation is controlled by a 
correlation parameter $\rho_1$, which modifies the orientation of the covariance 
structure while preserving its overall scale.

\end{description}

Combining these effects yields a wide range of scenarios, encompassing both pure and mixed regimes of dilatation, location shift, concentration, and deformation.

\begin{table}[ht]
\centering
\caption{Simulation settings for dilatation, location shift, concentration, and deformation regimes \label{tab:contscentransf}}
\begin{tabular}{lccc}
\hline
{Scenario} & $k$ (shift) & $\ell$ (scale) & $\rho_1$ (dependence) \\
\hline
Dilatation & 0   & (2, 10, 100)   & $\rho_0$ \\
Shift  & (5, 10, 50)   & 1     & $\rho_0$ \\
Concentration & 0   & (0.5, 0.1, 0.01)   & $\rho_0$ \\
Deformation & 0   & 1     & (0.5, 0.7, 0.95) \\
\hline
\end{tabular}
\end{table}

\paragraph{Generation of outliers.}

For each contamination level $r$, a single random permutation $\pi$ of $\{1,\dots,n\}$ is drawn once, and the outlier set is defined as $\mathcal{I}_r = \{\pi(1), \dots, \pi(n_r)\}$ with $n_r = \lfloor rn \rfloor$ the number of outliers. Consequently, observations in $\mathcal{I}_r$ are then sampled from $\mathcal{F}_1$, and the remaining ones from $\mathcal{F}_0$.
Since the permutations are nested across increasing values of $r$, the 
resulting outlier sets are nested as well, ensuring a coherent and 
variance-reduced comparison between contamination levels.

\paragraph{Evaluation criteria.}\label{par:evalcriteria}
The accuracy of covariance matrix estimation is assessed by means of the Frobenius norm errors, whereas the performance of outlier detection is evaluated through four complementary measures:
the false negative rate (FNR),
the false positive rate (FPR),
the area under the receiver operating characteristic curve (AUC), and
the classification accuracy (ACC), defined as the proportion of correctly 
classified entries of the estimated precision matrix, i.e., $1 - (\text{number of off-diagonal misclassifications}) / 
(\text{total number of off-diagonal entries})$.
Computation times are also recorded. The sample covariance online estimator is not reported in the computation time comparison, as no reference implementation is available to our knowledge.

\subsection{Results} 

We separate the results into two groups of contamination scenarios that induce qualitatively different distortions: the \emph{dilatation} and \emph{shift} (decentering) settings, in which the contaminated population is displaced or rescaled away from the central mass, yielding a larger separation between the two populations, and the \emph{concentration} and \emph{deformation} settings, in which the contaminated observations remain close to the central structure of the data.
The performance of the evaluation criteria is investigated in each case with respect to variations in the contamination level and the adaptation parameter.

\begin{figure}[!ht]
\centering

\textbf{Dilatation} \\[1em]

\begin{tabular}{c c}

&
\begin{tabular*}{0.95\textwidth}{@{\extracolsep{\fill}}ccccc@{}}
\textbf{\hspace{0.3cm}\shortstack{Frobenius \\ norm error}} &
\textbf{FNR} &
\textbf{FPR} &
\textbf{AUC\hspace{0.5cm}} &
\textbf{ACC \hspace{0.5cm}}
\end{tabular*}
\\[0.5em]
\\[0.5em]
\raisebox{1.5\height}{\rotatebox{90}{$\ell = 2$}}
&
\includegraphics[
width=0.95\textwidth,
height=4cm
]{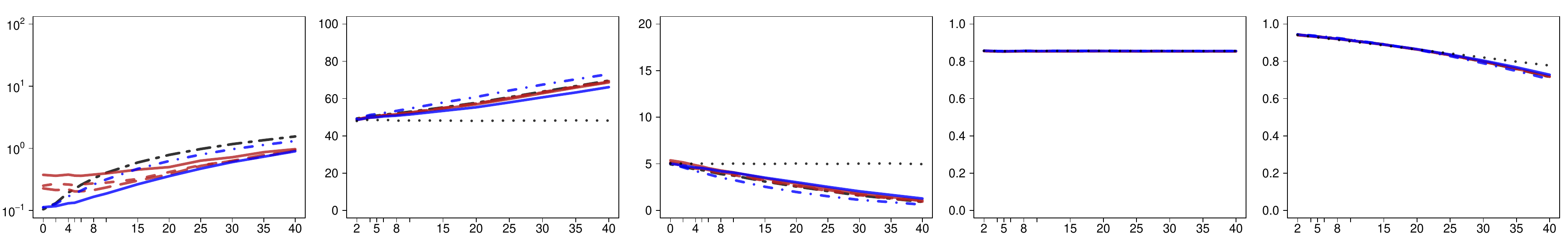}
\\[1em]

 \raisebox{1.5\height}{\rotatebox{90}{$\ell = 10$}}
&
\includegraphics[
width=0.95\textwidth,
height=4cm
]{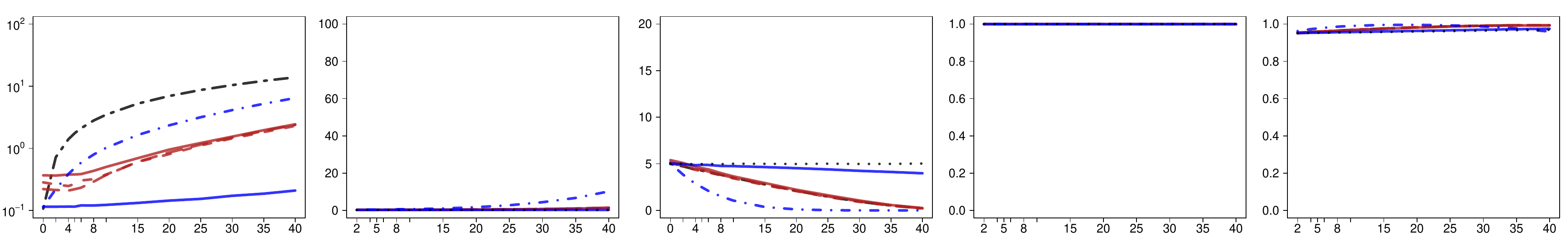}
\\[1em]

\raisebox{1.25\height}{\rotatebox{90}{$\ell = 100$}}
&
\includegraphics[
width=0.95\textwidth,
height=4cm
]{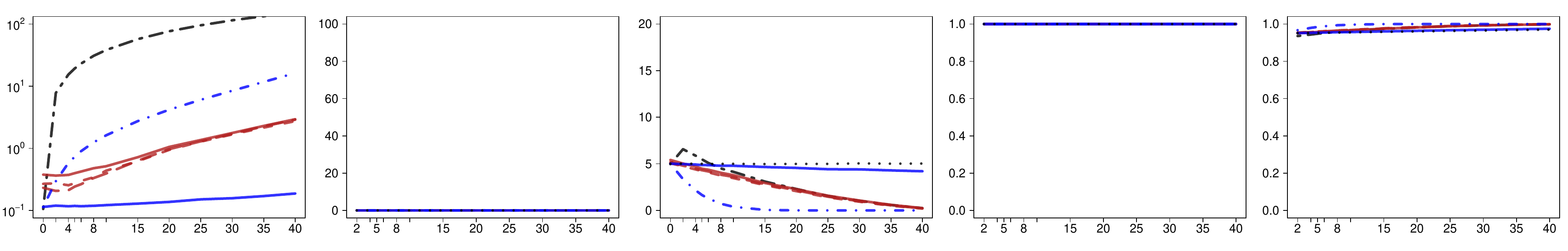}
\\

\end{tabular}

\caption{
  Evaluation criteria, Frobenius norm error, FNR, FPR, ACC and AUC as a function of the contamination rate for the \textbf{dilatation} scenario, with $\ell \in (2, 10, 100)$. \\
  Legend: 
\legendmethods}\label{fig:dilatation}
\end{figure}

\paragraph{Dilatation scenarios.}
Figure~\ref{fig:dilatation} presents the performance of the different estimation 
methods under the dilatation scenario.
Scale inflation exerts a far more detrimental effect on the empirical covariance estimator, as it directly incorporates the second-order contribution of contaminated observations; in contrast, the robust alternatives remain stable.
Among the robust methods, \textit{MCD} performs better under scale inflation, whereas the proposed methods outperform \textit{MCD} when the contaminating distribution 
exhibits a smaller covariance structure. Detection nonetheless remains effective: the inflated variance moves the contaminated observations away from the main data cluster in Mahalanobis distance, ensuring the two populations remain well separated despite the distortion of the estimated covariance.

\begin{figure}[!ht]
\centering

\textbf{Location shift} \\[1em]

\begin{tabular}{c c}
&
\begin{tabular*}{0.95\textwidth}{@{\extracolsep{\fill}}ccccc@{}}
\textbf{\hspace{0.3cm}\shortstack{Frobenius \\ norm error}} &
\textbf{FNR} &
\textbf{FPR} &
\textbf{AUC \hspace{0.5cm}} &
\textbf{ACC \hspace{0.5cm}}
\end{tabular*}
\\[1em]
\raisebox{1.5\height}{\rotatebox{90}{$k = 5$}}
&
\includegraphics[
width=0.95\textwidth,
height=4cm
]{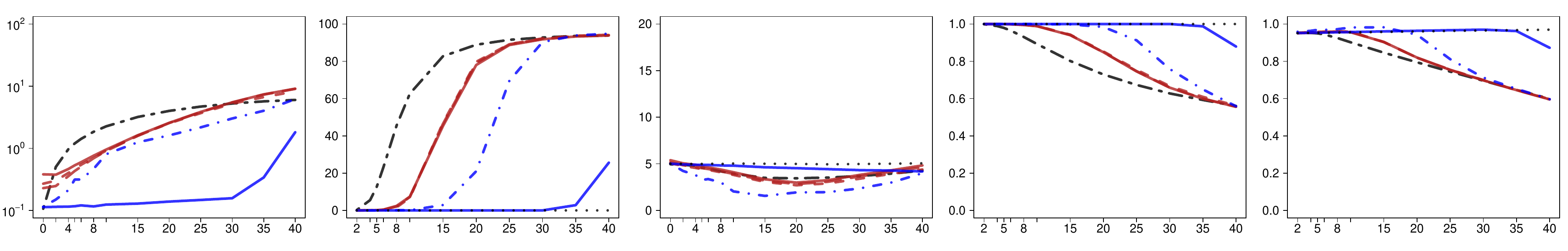}
\\[1em]

\raisebox{1.5\height}{\rotatebox{90}{$k = 10$}}
&
\includegraphics[
width=0.95\textwidth,
height=4cm
]{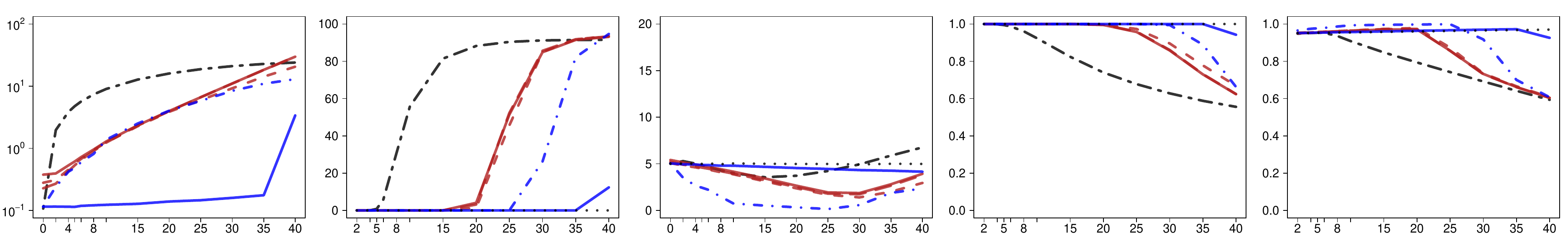}
\\[1em]

\raisebox{1.25\height}{\rotatebox{90}{$k = 50$}}
&
\includegraphics[
width=0.95\textwidth,
height=4cm
]{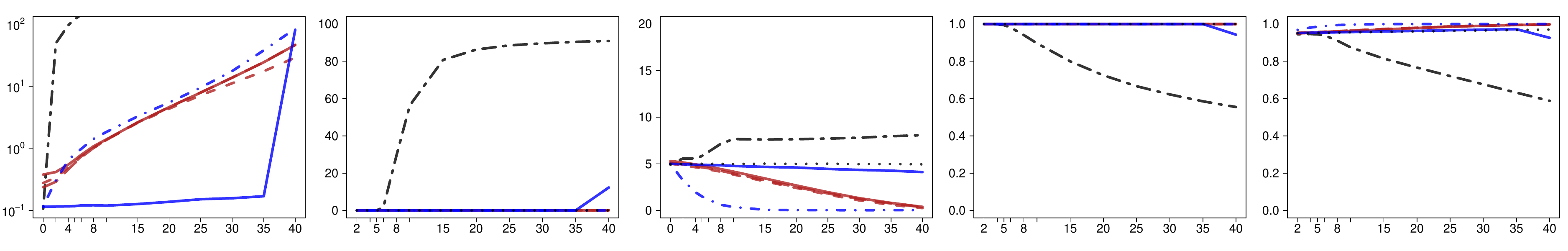}
\\
\end{tabular}

\caption{
  Evaluation criteria, Frobenius norm error, FNR, FPR, ACC and AUC as a function of the contamination rate for the \textbf{location shift} scenario, with increasing values of the shift parameter $k$. \textit{Legend: same as Figure~\ref{fig:dilatation}.}
}
\label{fig:shift}
\end{figure}

\paragraph{Location shift scenarios}
Figure~\ref{fig:shift} presents the performance of the different estimation 
methods under the shift scenario. The robust methods maintain a stable estimation of the covariance matrix throughout the simulation, 
whereas the empirical covariance estimator severely degrades even at moderate 
contamination levels. However, the detection behavior differs from that of
dilatation scenario: the false positive rate (FPR) decreases rapidly because the contaminated 
observations move far from the bulk. This creates a stricter detection rule, which in turn flags fewer observations as outliers. 
Despite this, detection remains effective up to approximately $35\%$ contamination for $k = 50$, and the 
increased separation between the populations yields high classification accuracy. 
This behavior occurs because, under a pure shift, the central 
location moves while the covariance structure of the bulk is preserved; consequently, the 
inlier distances stay essentially unchanged and only the outlier scores are 
pushed upward.

\begin{figure}[!ht]
\centering

\textbf{Concentration} \\[1em]

\begin{tabular}{c c}
&
\begin{tabular*}{0.95\textwidth}{@{\extracolsep{\fill}}ccccc@{}}
\textbf{\hspace{0.3cm}\shortstack{Frobenius \\ norm error}} &
\textbf{FNR} &
\textbf{FPR} &
\textbf{AUC \hspace{0.5cm}} &
\textbf{ACC \hspace{0.5cm}}
\end{tabular*}
\\[1em]
\raisebox{1.5\height}{\rotatebox{90}{$\ell = 0.5$}}
&
\includegraphics[
width=0.95\textwidth,
height=4cm
]{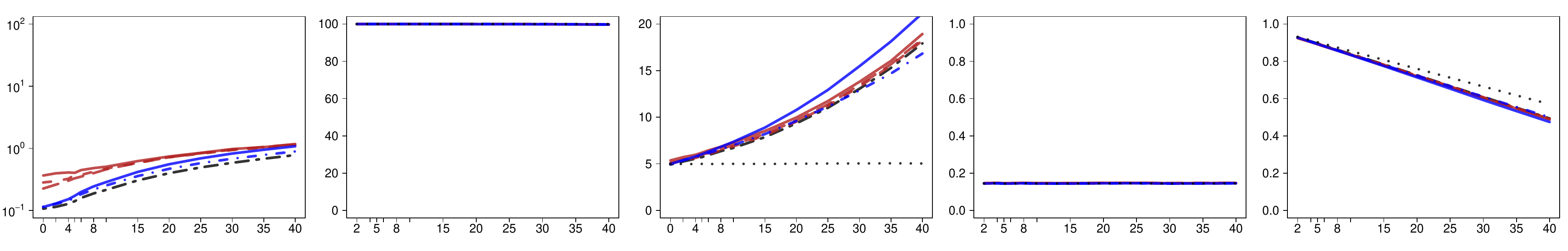}
\\[1em]

\raisebox{1.5\height}{\rotatebox{90}{$\ell = 0.1$}}
&
\includegraphics[
width=0.95\textwidth,
height=4cm
]{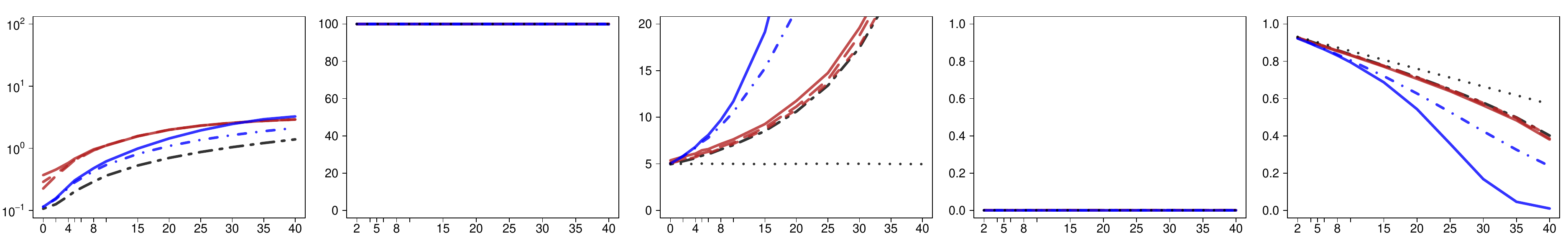}
\\[1em]

\raisebox{1.25\height}{\rotatebox{90}{$\ell = 0.01$}} 
&
\includegraphics[
width=0.95\textwidth,
height=4cm
]{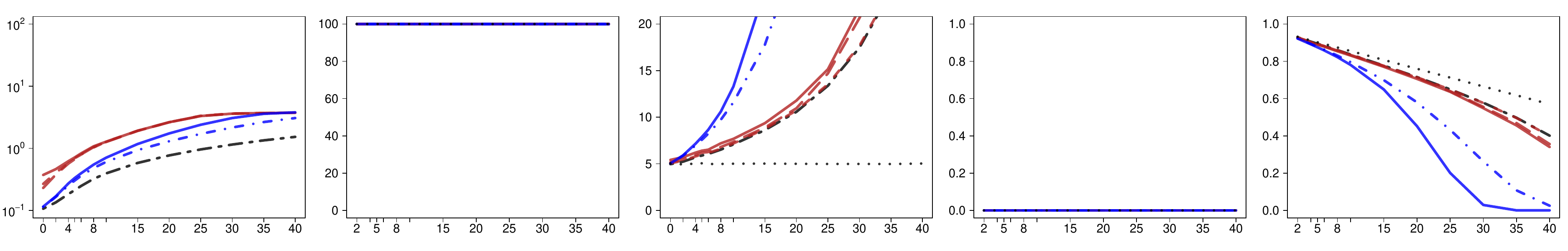}
\\
\end{tabular}

\vspace{0.8em}

\caption{
  Evaluation criteria, Frobenius norm error, FNR, FPR, ACC and AUC as a function of the contamination rate for the \textbf{concentration} scenario, with decreasing values of $\ell$. \textit{Legend: same as Figure~\ref{fig:dilatation}.
}}
\label{fig:concentration}
\end{figure}

\paragraph{Concentration scenarios.}
Figure~\ref{fig:concentration} presents the performance of the different estimation 
methods under the concentration scenario. Observe that the proposed methods maintain accurate covariance matrix estimations under concentration, whereas \textit{OGK} and \textit{MCD} degrade as contamination increases.
Under this regime, both ACC and AUC remain low because the
outliers lie close to the bulk, making discrimination inherently difficult.
Specifically, \textit{FAST-MCD} fails 
because its trimming mechanism assumes less than 50\% contamination and distant outliers; 
when these conditions are violated, it either discards inliers or retains outliers, thereby biasing the scatter estimate.
Similarly, \textit{OGK} suffers from a comparable bias when outliers are concentrated along directions correlated with the bulk, yielding an overly permissive threshold.
In contrast, the proposed online and streaming methods continuously adapt their thresholds to the empirical 
distribution of Mahalanobis distances. This keeps the decision boundary close to 
the true separation, resulting in fewer false discoveries than \textit{OGK} and \textit{MCD} despite an FPR above the nominal 5\% level.
Note that the deformation scenario, which shares 
the same qualitative behavior, is detailed in  Appendix~\ref{app:deformation}.

\begin{figure}[!ht]
    \centering
    \includegraphics[width=\textwidth]{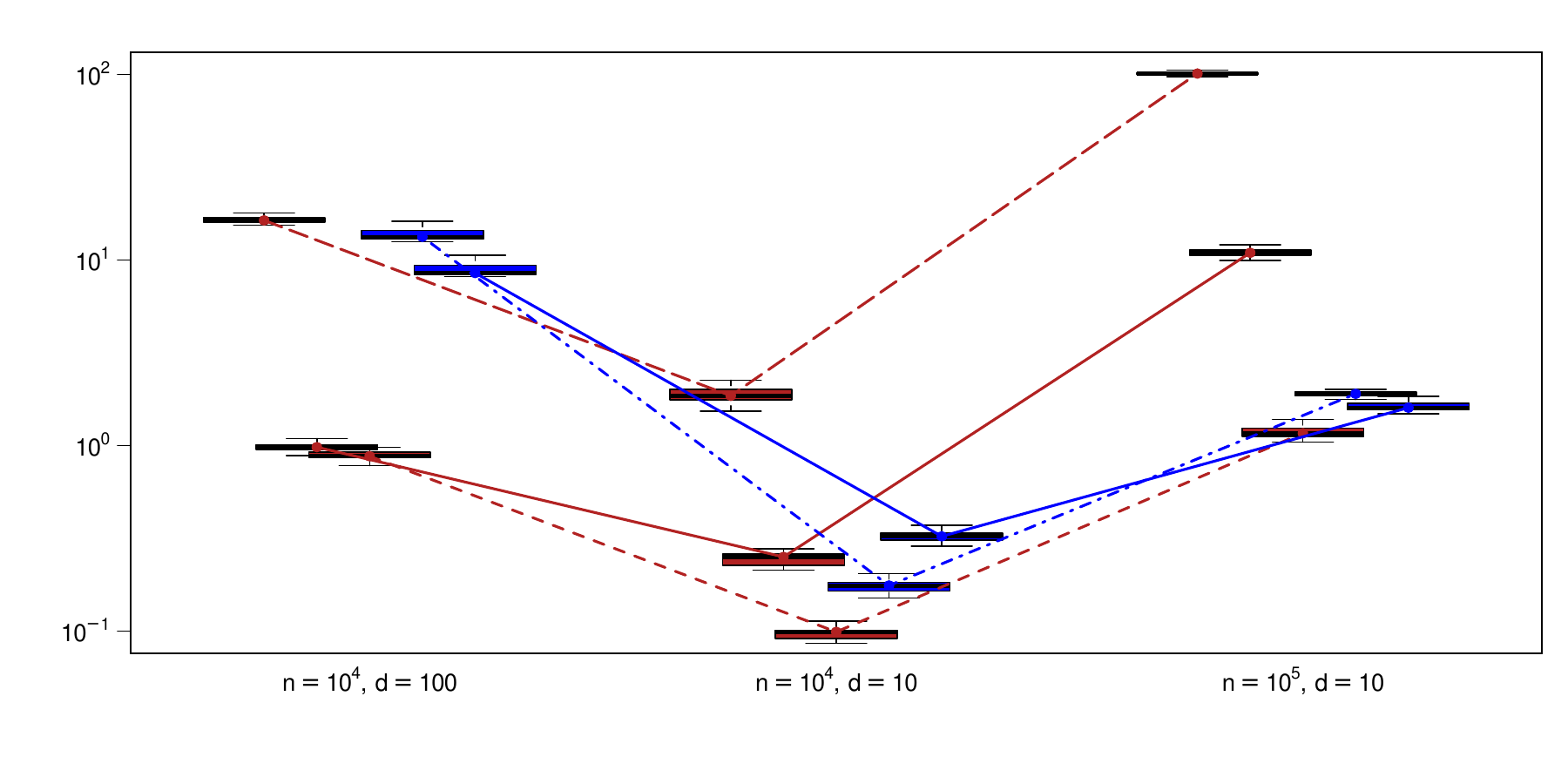}
    \caption{
        Computation time for different $(n,d)$ configurations. \textit{Legend: same as Figure~\ref{fig:dilatation}.}
    }
    \label{fig:computation_times}
\end{figure}

\paragraph{Computation times.}
The proposed online and streaming methods perform similarly across all configurations, their main difference being computational cost. Figure~\ref{fig:computation_times} shows that streaming is consistently faster, as it processes observations in batches of size $10$ and thus requires fewer matrix diagonalisations. The gain is largest for $(n,d)=(10^4,100)$ and smaller for $(10^5,10)$, where the increased number of batches reduces the relative advantage. Computation times nonetheless remain comparable to those of the benchmark methods considered. The computational benefit of streaming therefore stems mainly from the reduction in the number of matrix diagonalisation operations.

\paragraph{Performance parity and practical advantage of the online procedure.}
The online and streaming procedures achieve comparable statistical performance 
to their offline counterpart, confirming that the incremental formulation does 
not degrade the quality of the estimated median covariation matrix. Unlike the 
offline procedure, which requires the full dataset before producing an estimate 
and can only deliver a decision once all observations have been collected, the 
online procedure updates its estimate and issues a detection decision 
instantaneously as each observation arrives. This makes it the preferred choice 
for real-time streaming applications, where waiting for the complete data to 
become available is not an option.

\subsection{Real data experiments}

We now illustrate the application of the proposed online methods to real-world data.

\paragraph{Description of the dataset.}
We evaluate our methods on the Server Machine Dataset (SMD), a widely used benchmark for multivariate time-series anomaly detection in streaming settings~\cite{su2019robust}.

The publicly available Server Machine Dataset (SMD) available in Kaggle\citep{kaggle_smd} comprises multiple sensors and system-level metrics collected at one-minute intervals across 28 server machines. Each observation is a multivariate vector capturing real-time machine activity, including CPU utilization, memory consumption, network traffic, and other operational statistics.

Our experiments utilize sequences of approximately 23,000 to 28,000 observations across 38 variables. The data exhibit pronounced temporal dependence, strong correlations, and substantial redundancy among the monitored variables. This high collinearity frequently renders the corresponding covariance matrices ill-conditioned or nearly singular, creating a challenging environment for robust online covariance estimation and anomaly detection.
The considered test sequence also contains anomalous observations that are explicitly labeled in the dataset. More precisely, each timestamp is associated with a binary label, with 1 indicating an anomalous observation and 0 indicating normal operating conditions. These labels are used exclusively for performance assessment and threshold calibration and are not incorporated into the online estimation procedure. The SMD dataset therefore provides a realistic and challenging benchmark for evaluating robust online estimation procedures and streaming outlier detection methods under temporally dependent, highly correlated, and potentially ill-conditioned data.

\paragraph{Data preprocessing.}
Prior to the analysis, the SMD data are preprocessed consistently across all 
machines to improve numerical stability and ensure comparability across methods. 
The detailed pipeline — including variable filtering, redundancy removal, robust 
standardization, and the batch size used by the streaming procedures — is 
described in Appendix~\ref{app:preprocessing}.

\begin{figure}[ht!]
    \centering

    \begin{subfigure}[b]{0.32\textwidth}
        \centering
        \textbf{Machine 5}
    \end{subfigure}
    \hfill
    \begin{subfigure}[b]{0.32\textwidth}
        \centering
        \textbf{Machine 33}
    \end{subfigure}
    \hfill
    \begin{subfigure}[b]{0.32\textwidth}
        \centering
        \textbf{Machine 56}
    \end{subfigure}

    \vspace{0.5em}

    \begin{subfigure}[b]{0.32\textwidth}
        \centering
        \includegraphics[width=\textwidth]{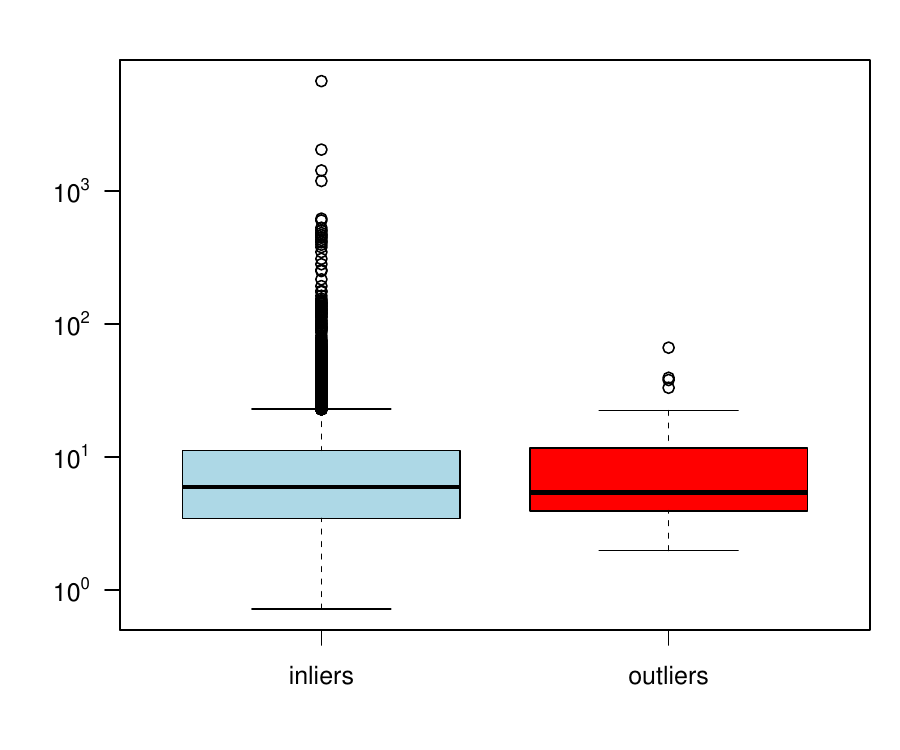}
    \end{subfigure}
    \hfill
    \begin{subfigure}[b]{0.32\textwidth}
        \centering
        \includegraphics[width=\textwidth]{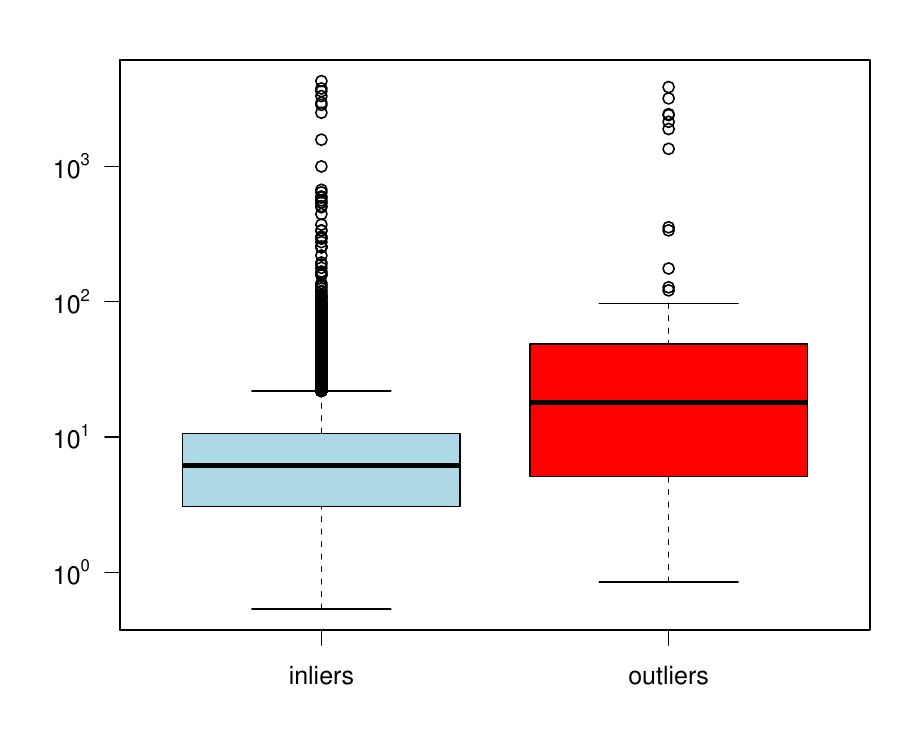}
    \end{subfigure}
    \hfill
    \begin{subfigure}[b]{0.32\textwidth}
        \centering
        \includegraphics[width=\textwidth]{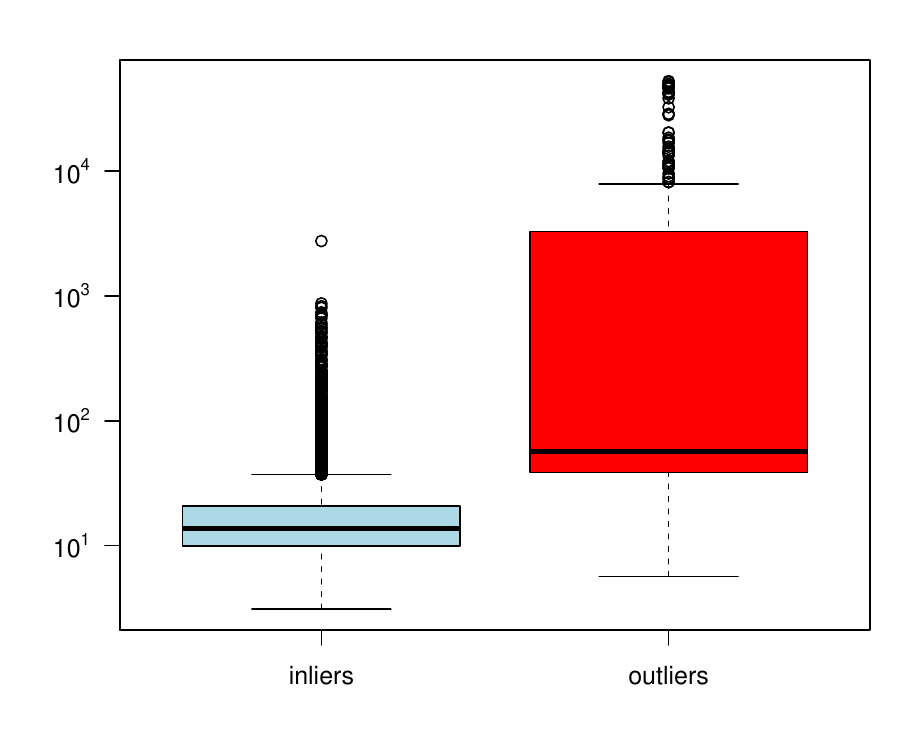}
    \end{subfigure}

    \vspace{1em}

    \begin{subfigure}[b]{0.32\textwidth}
        \centering
        \includegraphics[width=\textwidth]{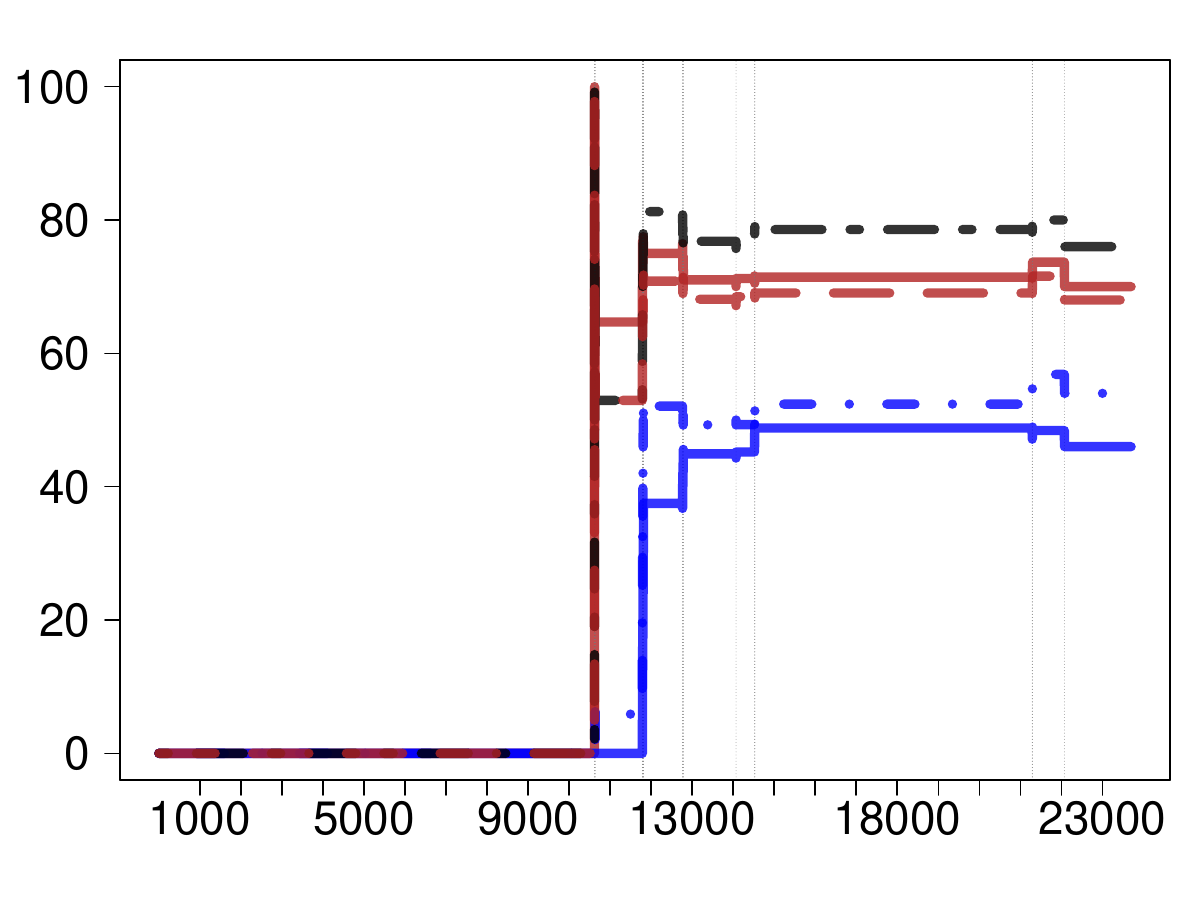}
    \end{subfigure}
    \hfill
    \begin{subfigure}[b]{0.32\textwidth}
        \centering
        \includegraphics[width=\textwidth]{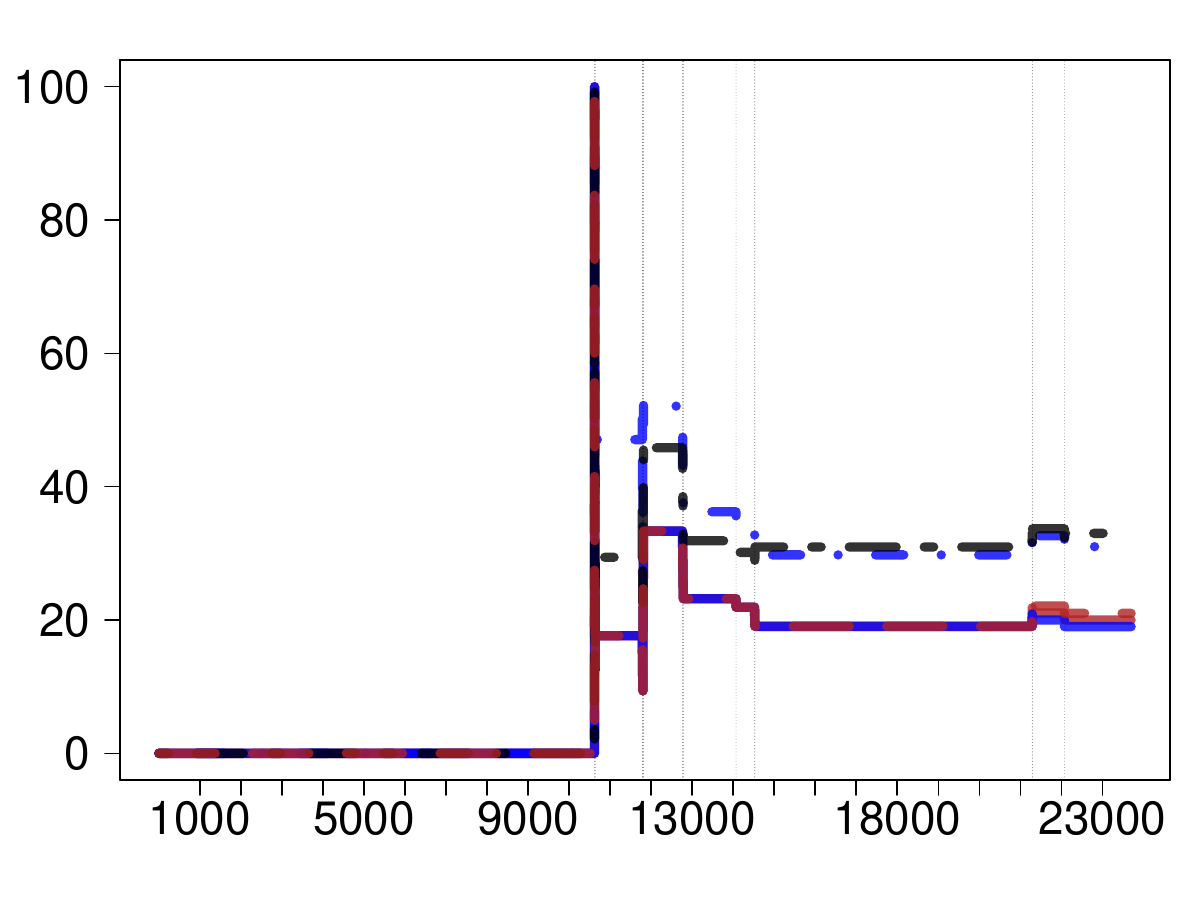}
    \end{subfigure}
    \hfill
    \begin{subfigure}[b]{0.32\textwidth}
        \centering
        \includegraphics[width=\textwidth]{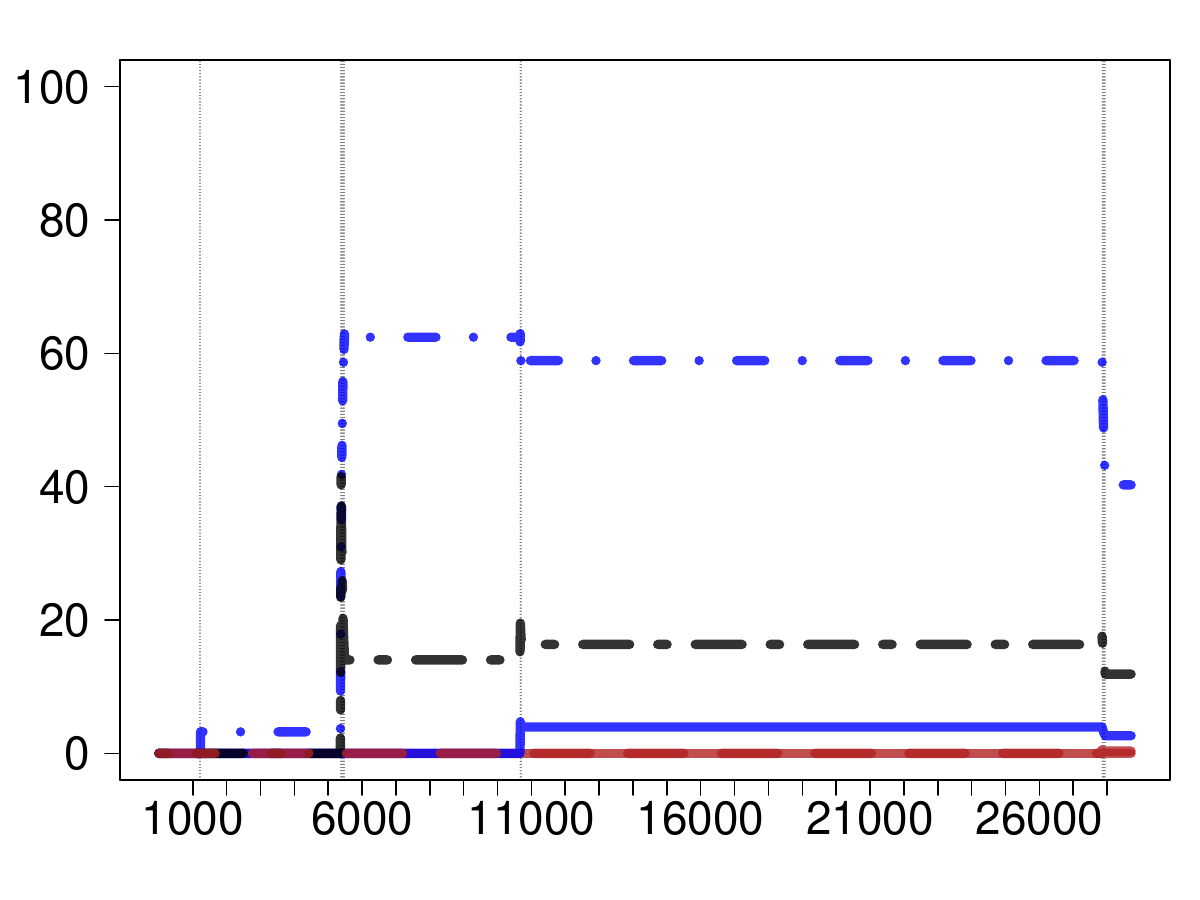}
    \end{subfigure}

    \vspace{1em}

    \begin{subfigure}[b]{0.32\textwidth}
        \centering
        \includegraphics[width=\textwidth]{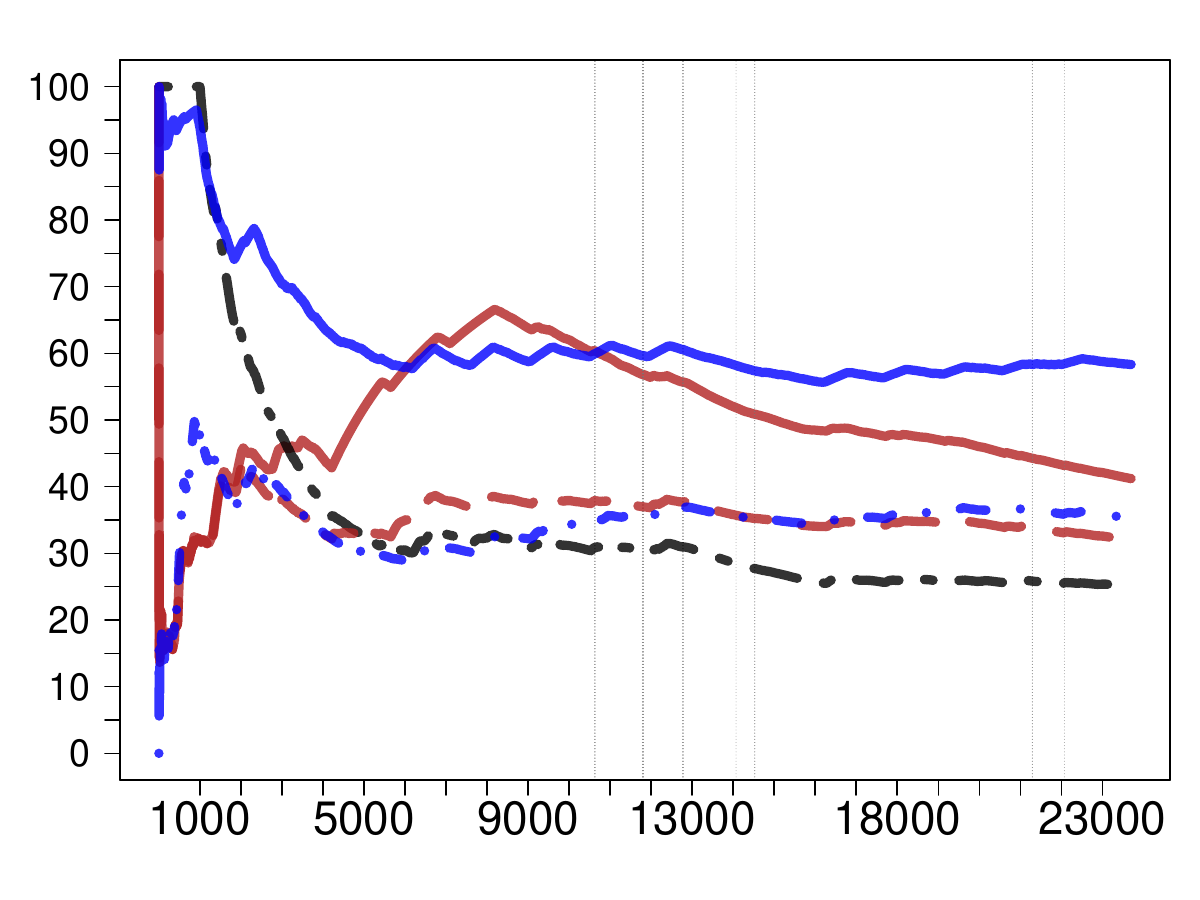}
    \end{subfigure}
    \hfill
    \begin{subfigure}[b]{0.32\textwidth}
        \centering
        \includegraphics[width=\textwidth]{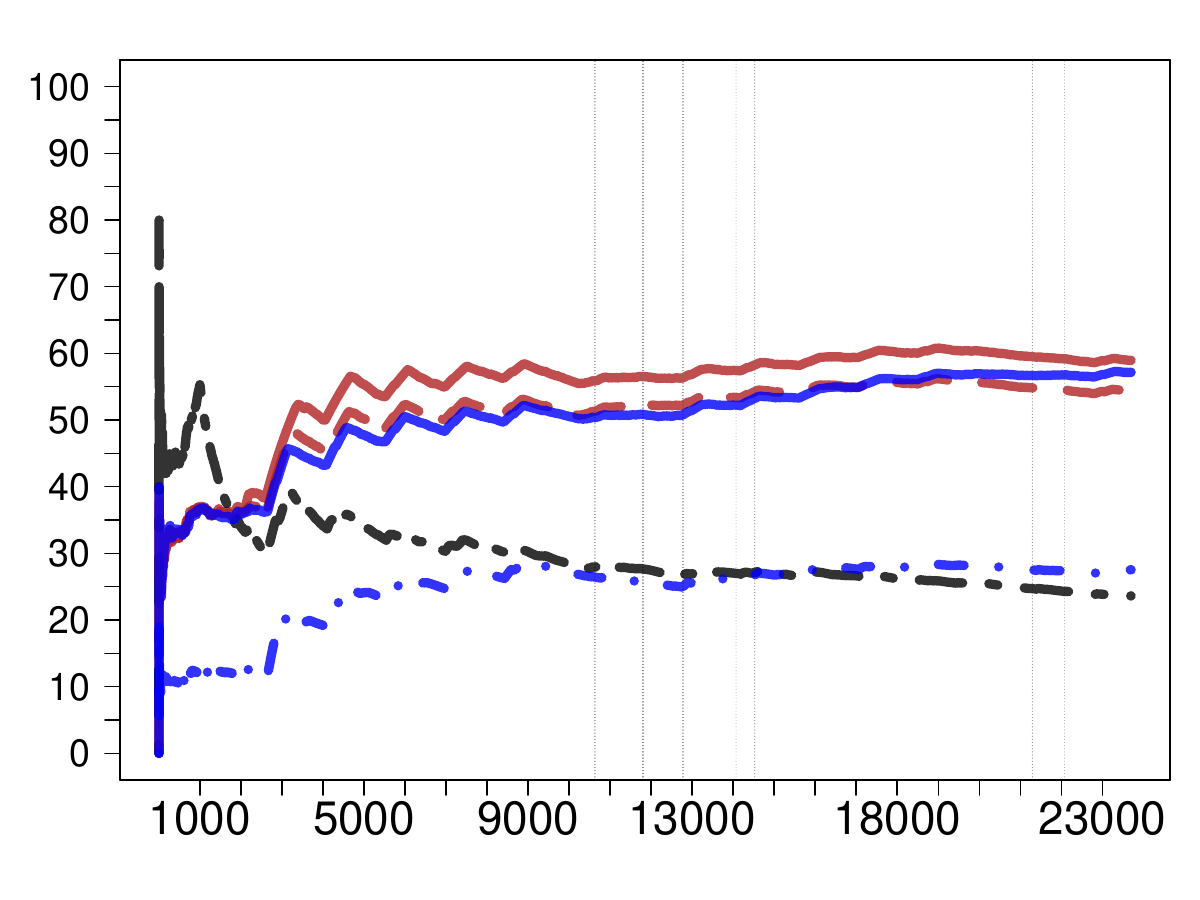}
    \end{subfigure}
    \hfill
    \begin{subfigure}[b]{0.32\textwidth}
        \centering
        \includegraphics[width=\textwidth]{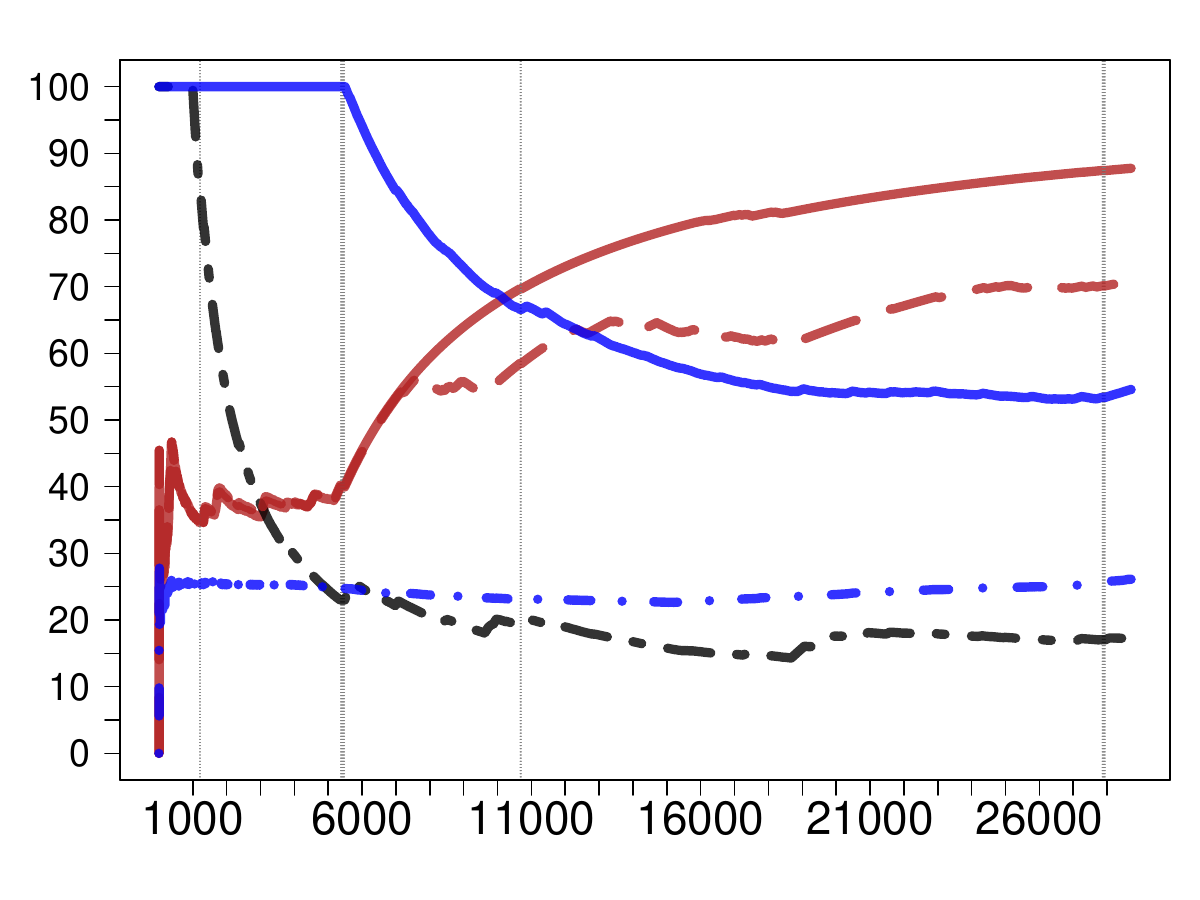}
    \end{subfigure}

    \vspace{1em}

    \begin{subfigure}[b]{0.32\textwidth}
        \centering
        \includegraphics[width=\textwidth]{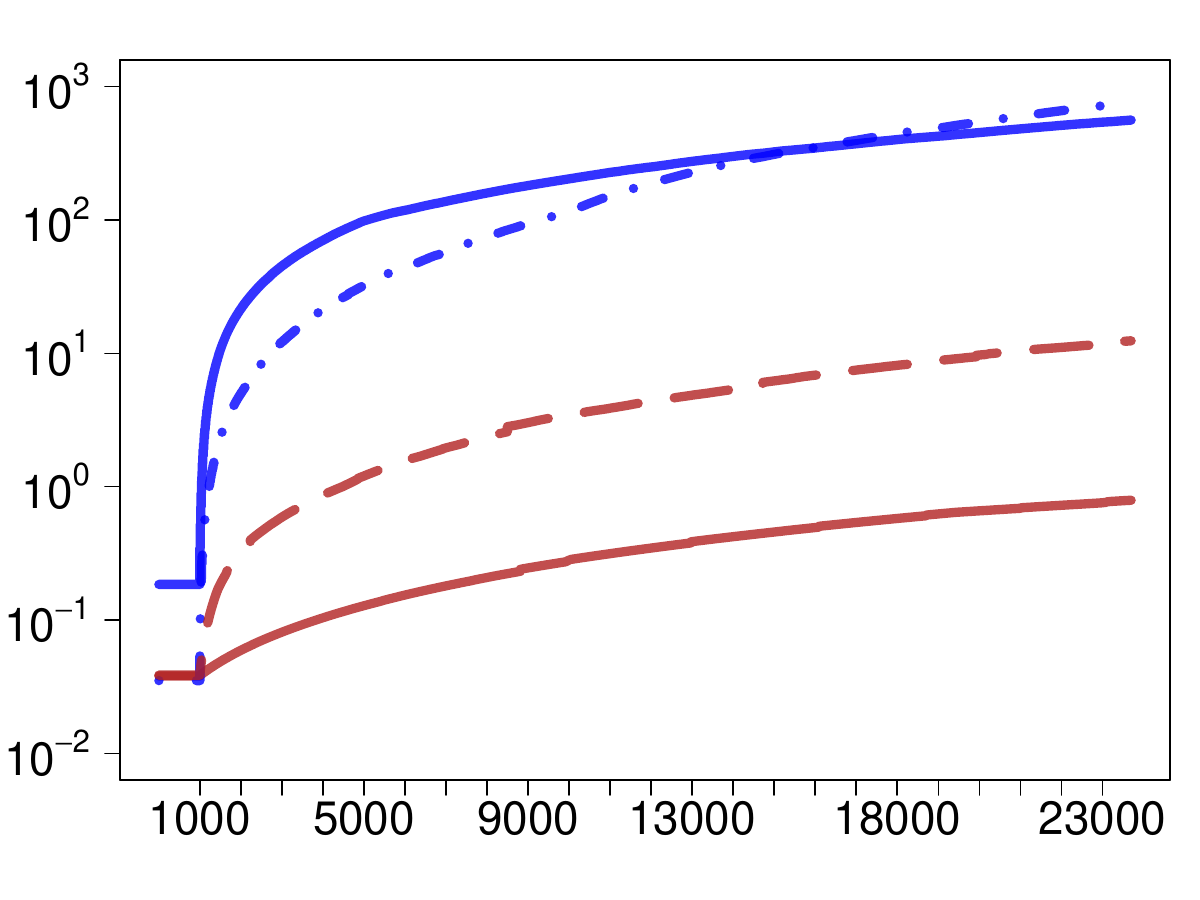}
    \end{subfigure}
    \hfill
    \begin{subfigure}[b]{0.32\textwidth}
        \centering
        \includegraphics[width=\textwidth]{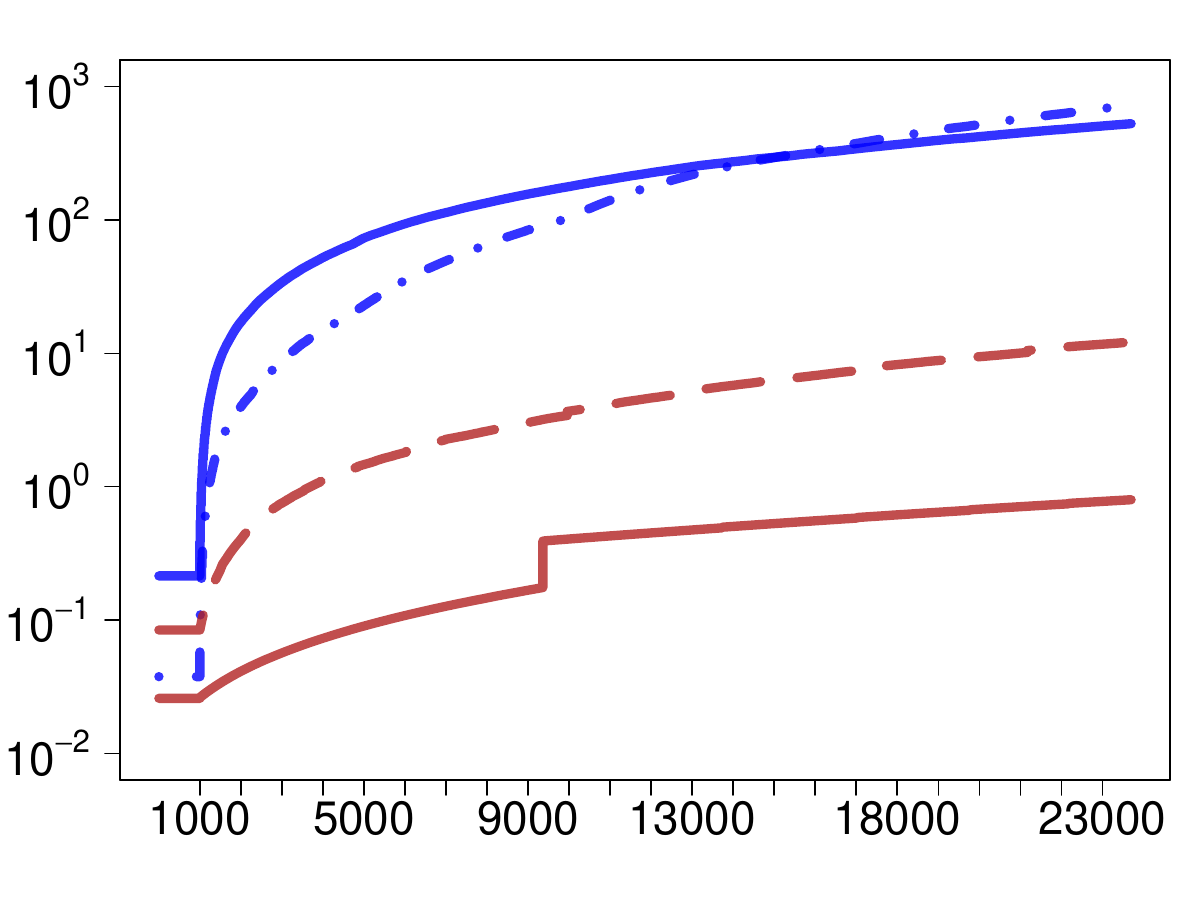}
    \end{subfigure}
    \hfill
    \begin{subfigure}[b]{0.32\textwidth}
        \centering
        \includegraphics[width=\textwidth]{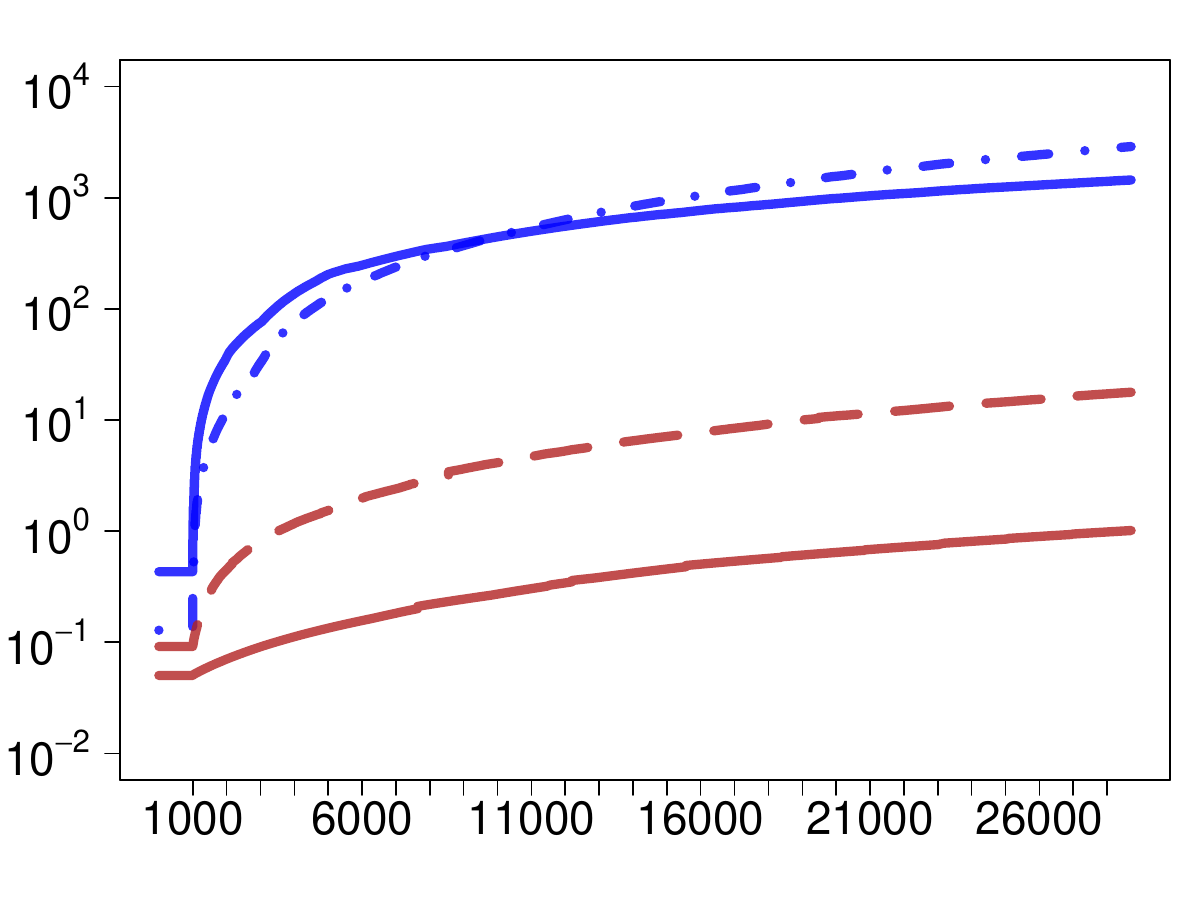}
    \end{subfigure}

    \caption{Trajectories for machines 5, 33, and 56.
    The top row reports the boxplots of the squared Mahalanobis distances
    for inliers and outliers; subsequent rows report
    respectively the false negative rate (FNR), the false positive rate (FPR) and the
    cumulative computation times as a function of the number of processed
    observations. Vertical grey lines indicate the positions of the true
    anomalies.  \textit{Legend: same as Figure~\ref{fig:dilatation}.} 
  }
    \label{fig:trajsmd}
\end{figure}

\paragraph{Objective.}
We evaluate anomaly detection performance in both streaming and batch settings on the SMD dataset by jointly analyzing cumulative false positive and false negative rates.
To highlight the problem's intrinsic difficulty, Figure~\ref{fig:trajsmd} (top row) presents boxplots of the squared Mahalanobis distances for machines 5, 33, and 56. These distances are computed using the empirical mean and covariance matrix derived from uncontaminated data.

\paragraph{False negative trajectories.}
The boxplots of inlier and outlier distances reveal substantial overlap on the 
three machines, confirming that detection under the Mahalanobis criterion is 
intrinsically difficult. Despite this, the online and streaming procedures 
rapidly reduce their false negative rate as observations accumulate, reaching 
low levels within the first few thousand points and remaining stable thereafter. 
In contrast, \textit{OGK} and \textit{MCD} require substantially more observations to converge, 
and their FNR remains markedly higher throughout the stream on machines 33 and 
56. This gap is further amplified in configurations where the covariance 
matrices exhibit an almost rank-one structure: in such regimes, MCD struggles 
to identify a well-conditioned $h$-subset, since the determinant of the 
estimated scatter matrix becomes extremely sensitive to the subset selection, 
which destabilises the resulting detection rule and inflates the false negative 
rate. This illustrates a key advantage of the proposed online formulation: it 
detects anomalies as soon as they arrive, without waiting for a sufficient 
history to produce a reliable estimate.

\paragraph{False positive trajectories.}
The false positive rates confirm the same pattern from the opposite side. The 
online and streaming procedures maintain controlled FPR values across the three 
machines, whereas the empirical covariance estimator degrades sharply on machine 
56 and the \textit{MCD} estimator exhibits pronounced fluctuations on machines 5 and 33. 
\textit{OGK} remains conservative but at the cost of a high false negative rate, and 
therefore offers an unfavourable trade-off. The online procedure benefits from 
its adaptive threshold, which is continuously recalibrated against the empirical 
distribution of the observed distances, keeping the false positive rate within 
acceptable bounds despite the overlap between the two populations.

\paragraph{Cumulative computation times.}
The cumulative computation times reveal the most striking practical advantage of 
the proposed methods. The streaming procedure is consistently the fastest, 
followed by the online procedure, whereas \textit{OGK} and \textit{MCD} exhibit a much 
steeper growth, their cost being dominated by repeated pairwise medians and 
iterative C-steps. As in the simulated data experiments, the streaming procedure benefits 
from a reduction in the number of matrix diagonalisation operations compared 
with the purely online procedure, since it processes the observations in batches 
rather than one at a time. The cumulative running time is reported successively over the 
observation ranges $(1{:}n_0)$, $(1{:}(n_0+\mathrm{batch}))$, $(1{:}(n_0+2\,\mathrm{batch}))$, and so on, so that each point represents the 
total computational time accumulated from the beginning of the experiment up to  the corresponding number of processed observations. For all machines for which the streaming computations could 
be completed, the total accumulated time remained lower than the time required to process a single machine using either \textit{OGK} or \textit{MCD}. This gap widens as the stream progresses, confirming that the online and streaming formulations are 
not only statistically competitive but also operationally viable for real-time anomaly detection on long data streams.

\paragraph{Concluding remarks.}
To conclude, the proposed online and 
streaming estimators achieve detection performance comparable to their offline 
counterpart while operating under a strict streaming constraint, which is not 
the case for any of the robust competitors considered. Moreover, they remain 
computationally viable on long data streams, whereas \textit{FAST-MCD} and \textit{OGK} are 
intrinsically batch procedures: their estimates depend on global quantities 
that cannot be updated incrementally, and each new observation would in  principle require recomputing the pairwise medians or the optimal $h$-subset from scratch. This structural barrier, more than the raw computational cost, 
explains the absence of online implementations of these estimators and justifies the approach developed here.

\section*{Declaration of generative AI and AI-assisted technologies in the manuscript preparation process}

During the preparation of this work, the authors used ChatGPT and DeepSeek (free versions) to improve the language, readability, and formatting of the literature review section and the figures. No AI tools were used in the mathematical formulations or methodological and algorithmic developments. After using these tools, the authors reviewed, verified, and edited the content as needed, and take full responsibility for the content of the published article.

\bibliographystyle{apalike}
\bibliography{ref} 

@article{pearson1901liii,
  title={LIII. On lines and planes of closest fit to systems of points in space},
  author={Pearson, Karl},
  journal={The London, Edinburgh, and Dublin philosophical magazine and journal of science},
  volume={2},
  number={11},
  pages={559--572},
  year={1901},
  publisher={Taylor \& Francis}
}

@article{hotelling1931generalization,
  title={The generalization of Student's ratio},
  author={Hotelling, Harold and others},
  year={1931},
  journal = {Annals of Mathematical Statistics},
  publisher={Springer}
}

@article{hotelling1933analysis,
  title={Analysis of a complex of statistical variables into principal components.},
  author={Hotelling, Harold},
  journal={Journal of educational psychology},
  volume={24},
  number={6},
  pages={417},
  year={1933},
  publisher={Warwick \& York}
}

@article{weiszfeld1937point,
  title={On the point for which the sum of the distances to n given points is
 minimum},
  author={Weiszfeld, Endre},
  journal={Tohoku Mathematical Journal, First Series},
  volume={43},
  pages={355--386},
  year={1937},
  publisher={Mathematical Institute, Tohoku University}
}

@article{robbins1951stochastic,
  title={A Stochastic Approximation Method},
  author={Robbins, Herbert and Monro, Sutton},
  journal={The Annals of Mathematical Statistics},
  volume={22},
  number={3},
  pages={400--407},
  year={1951}
}

@article{projfriedmanTukey1974,
author = {Friedman, Jerome and Tukey, J.W.},
year = {1974},
month = {10},
pages = {881 - 890},
title = {A Projection Pursuit Algorithm for Exploratory Data Analysis,” IEEE Transactions on Computers, C-23, 881-889},
volume = {C 23},
journal = {Computers, IEEE Transactions on},
doi = {10.1109/T-C.1974.224051}
}

@book{stahel1981breakdown,
  title={Breakdown of covariance estimators},
  author={Stahel, Werner A},
  year={1981},
  publisher={Fachgruppe f{\"u}r Statistik, Eidgen{\"o}ssische Techn. Hochsch.}
}

@techreport{donoho1982breakdown,
  title={Breakdown properties of multivariate location estimators},
  author={Donoho, David L},
  year={1982},
  institution={Technical report, Harvard University, Boston. URL http://www-stat. stanford~…}
}

@article{pena2001multivariate,
  title={Multivariate outlier detection and robust covariance matrix estimation},
  author={Pe{\~n}a, Daniel and Prieto, Francisco J},
  journal={Technometrics},
  volume={43},
  number={3},
  pages={286--310},
  year={2001},
  publisher={Taylor \& Francis}
}

@phdthesis{labopin2016methodes,
  title={M{\'e}thodes statistiques et d’apprentissage pour l’estimation de quantiles et de superquantiles dans des mod{\`e}les de codes num{\'e}riques ou stochastiques},
  author={Labopin-Richard, Tatiana},
  year={2016},
  school={Th{\`e}se de doctorat, Universit{\'e} Toulouse 3 Paul Sabatier}
}

@article{hampel1974influence,
  title={The influence curve and its role in robust estimation},
  author={Hampel, Frank R},
  journal={Journal of the american statistical association},
  volume={69},
  number={346},
  pages={383--393},
  year={1974},
  publisher={Taylor \& Francis}
}

@book{jolliffe1986mathematical,
  title={Mathematical and statistical properties of sample principal components},
  author={Jolliffe, Ian T},
  year={1986},
  publisher={Springer}
}

@article{kemperman1987median,
  title={The median of a finite measure on a Banach space},
  author={Kemperman, Johannes HB},
  journal={Statistical data analysis based on the L1-norm and related methods (Neuch{\^a}tel, 1987)},
  pages={217--230},
  year={1987}
}

@techreport{ruppert1988efficient,
  title={Efficient estimations from a slowly convergent Robbins-Monro process},
  author={Ruppert, David},
  year={1988},
  institution={Cornell University Operations Research and Industrial Engineering}
}

@article{rousseeuw1999fast,
  title={A fast algorithm for the minimum covariance determinant estimator},
  author={Rousseeuw, Peter J and Driessen, Katrien Van},
  journal={Technometrics},
  volume={41},
  number={3},
  pages={212--223},
  year={1999},
  publisher={Taylor \& Francis}
}

@article{vardi2000multivariate,
  title={The multivariate L 1-median and associated data depth},
  author={Vardi, Yehuda and Zhang, Cun-Hui},
  journal={Proceedings of the National Academy of Sciences},
  volume={97},
  number={4},
  pages={1423--1426},
  year={2000},
  publisher={The National Academy of Sciences}
}

@article{rousseeuw1985multivariate,
  title={Multivariate Estimation With High Breakdown Point},
  author={Rousseeuw, Peter},
  journal={Mathematical Statistics and Applications Vol. B},
  pages = {283--297},
  year={1985}
}

@article{rousseeuw1990unmasking,
  title={Unmasking multivariate outliers and leverage points},
  author={Rousseeuw, Peter J and Van Zomeren, Bert C},
  journal={Journal of the American Statistical association},
  volume={85},
  number={411},
  pages={633--639},
  year={1990},
  publisher={Taylor \& Francis}
}

@inproceedings{caussinus1990interesting,
  title={Interesting projections of multidimensional data by means of generalized principal component analyses},
  author={Caussinus, Henri and Ruiz, A},
  booktitle={Compstat: Proceedings in Computational Statistics, 9th Symposium held at Dubrovnik, Yugoslavia, 1990},
  pages={121--126},
  year={1990},
  organization={Springer}
}

@book{Jolliffe2002,
author = {Jolliffe, I.T.},
title= {Principal component analysis (2nd edition)},
year = {2002},
address = {Berlin},
publisher = {Springer Verlag},
}

@article{hubert2005robpca,
  title={ROBPCA: a new approach to robust principal component analysis},
  author={Hubert, Mia and Rousseeuw, Peter J and Vanden Branden, Karlien},
  journal={Technometrics},
  volume={47},
  number={1},
  pages={64--79},
  year={2005},
  publisher={Taylor \& Francis}
}

@article{tyler2009invariant,
  title={Invariant co-ordinate selection},
  author={Tyler, David E and Critchley, Frank and D{\"u}mbgen, Lutz and Oja, Hannu},
  journal={Journal of the Royal Statistical Society Series B: Statistical Methodology},
  volume={71},
  number={3},
  pages={549--592},
  year={2009},
  publisher={Oxford University Press}
}

@article{gervini2006robust,
  title={Robust Functional Data Analysis},
  author={Gervini, Daniel},
  journal={NSF Award Number 0604396. Directorate for Mathematical and Physical Sciences},
  volume={6},
  number={604396},
  pages={4396},
  year={2006}
}

@article{gervini2012outlier,
  title={Outlier detection and trimmed estimation for general functional data},
  author={Gervini, Daniel},
  journal={Statistica Sinica},
  pages={1639--1660},
  year={2012},
  publisher={JSTOR}
}

@article{beck2015weiszfeld,
  title={Weiszfeld’s method: Old and new results},
  author={Beck, Amir and Sabach, Shoham},
  journal={Journal of Optimization Theory and Applications},
  volume={164},
  pages={1--40},
  year={2015},
  publisher={Springer}
}

@article{godichon2024robust,
  title={A robust model-based clustering based on the geometric median and the median covariation matrix},
  author={Godichon-Baggioni, Antoine and Robin, St{\'e}phane},
  journal={Statistics and computing},
  volume={34},
  number={1},
  pages={55},
  year={2022},
  publisher={Springer}
}

@article{mokkadem2011generalization,
  title={A generalization of the averaging procedure: The use of two-time-scale algorithms},
  author={Mokkadem, Abdelkader and Pelletier, Mariane},
  journal={SIAM Journal on Control and Optimization},
  volume={49},
  number={4},
  pages={1523--1543},
  year={2011},
  publisher={SIAM}
}

@article{kraus2012dispersion,
  title={Dispersion operators and resistant second-order functional data analysis},
  author={Kraus, David and Panaretos, Victor M},
  journal={Biometrika},
  volume={99},
  number={4},
  pages={813--832},
  year={2012},
  publisher={Oxford University Press}
}

@article{cardot2017fast,
  title={Fast estimation of the median covariation matrix with application to online robust principal components analysis},
  author={Cardot, Herv{\'e} and Godichon-Baggioni, Antoine},
  journal={Test},
  volume={26},
  number={3},
  pages={461--480},
  year={2017},
  publisher={Springer}
}

@article{maronna2002robust,
  title={Robust estimates of location and dispersion for high-dimensional datasets},
  author={Maronna, Ricardo A and Zamar, Ruben H},
  journal={Technometrics},
  volume={44},
  number={4},
  pages={307--317},
  year={2002},
  publisher={Taylor \& Francis}
}

@article{cardot2013efficient,
  title={Efficient and fast estimation of the geometric median in Hilbert spaces with an averaged stochastic gradient algorithm},
  author={Cardot, Herv{\'e} and C{\'e}nac, Peggy and Zitt, Pierre-Andr{\'e}},
  journal = {Bernoulli},
  volume = {19},
 number = {1},
 pages = {18--43},
  year={2013}
}

@article{godichon2016estimating,
  title={Estimating the geometric median in Hilbert spaces with stochastic gradient algorithms: Lp and almost sure rates of convergence},
  author={Godichon-Baggioni, Antoine},
  journal={Journal of Multivariate Analysis},
  volume={146},
  pages={209--222},
  year={2016},
  publisher={Elsevier}
}

@article{godichon2023online,
  title={Online stochastic newton methods for estimating the geometric median and applications},
  author={Godichon-Baggioni, Antoine and Lu, Wei},
  journal={arXiv preprint arXiv:2304.00770},
  year={2023}
}

@article{boyer2023asymptotic,
  title={On the asymptotic rate of convergence of stochastic newton algorithms and their weighted averaged versions},
  author={Boyer, Claire and Godichon-Baggioni, Antoine},
  journal={Computational Optimization and Applications},
  volume={84},
  number={3},
  pages={921--972},
  year={2023},
  publisher={Springer}
}

@article{Godichon_Baggioni_2023,
   title={Non-asymptotic analysis of Stochastic approximation algorithms for streaming data},
   volume={27},
   ISSN={1262-3318},
   url={http://dx.doi.org/10.1051/ps/2023006},
   DOI={10.1051/ps/2023006},
   journal={ESAIM: Probability and Statistics},
   publisher={EDP Sciences},
   author={Godichon-Baggioni, Antoine and Werge, Nicklas and Wintenberger, Olivier},
   year={2023},
   pages={482–514} }

@article{haldane1948note,
  title={Note on the median of a multivariate distribution},
  author={Haldane, JBS},
  journal={Biometrika},
  volume={35},
  number={3-4},
  pages={414--417},
  year={1948},
  publisher={Oxford University Press}
}

@article{filzmoser2008outlier,
  title={Outlier identification in high dimensions},
  author={Filzmoser, Peter and Maronna, Ricardo and Werner, Mark},
  journal={Computational statistics \& data analysis},
  volume={52},
  number={3},
  pages={1694--1711},
  year={2008},
  publisher={Elsevier}
}

@article{gnanadesikan1972robust,
  title={Robust estimates, residuals, and outlier detection with multiresponse data},
  author={Gnanadesikan, Ramanathan and Kettenring, John R},
  journal={Biometrics},
  pages={81--124},
  year={1972},
  publisher={JSTOR}
}

@article{Cabana_2019,
  title={Multivariate outlier detection based on a robust Mahalanobis distance with shrinkage estimators},
  author={Cabana, Elisa and Lillo, Rosa E and Laniado, Henry},
  journal={Statistical papers},
  volume={62},
  pages={1583--1609},
  year={2021},
  publisher={Springer}
}

@Article{Ledoit_Wolf,
  author={Ledoit, Olivier and Wolf, Michael},
  title={{A well-conditioned estimator for large-dimensional covariance matrices}},
  journal={Journal of Multivariate Analysis},
  year=2004,
  volume={88},
  number={2},
  pages={365-411},
  month={February},
  doi={},
  url={https://ideas.repec.org/a/eee/jmvana/v88y2004i2p365-411.html}
}

@article{falkm,
  title={On mad and comedians},
  author={Falk, Michael},
  journal={Annals of the Institute of Statistical Mathematics},
  volume={49},
  pages={615--644},
  year={1997},
  publisher={Springer}
}

@inproceedings{su2019robust,
  author    = {Su, Ya and Zhao, Youjian and Niu, Chenhao and Liu, Rong 
               and Sun, Wei and Pei, Dan},
  title     = {Robust Anomaly Detection for Multivariate Time Series 
               through Stochastic Recurrent Neural Network},
  booktitle = {Proceedings of the 25th ACM SIGKDD International Conference 
               on Knowledge Discovery \& Data Mining (KDD '19)},
  pages     = {2828--2837},
  year      = {2019},
  publisher = {ACM},
  address   = {New York, NY, USA},
  doi       = {10.1145/3292500.3330672}
}

@misc{kaggle_smd,
  author       = {Mustafa G{\"u}sal},
  title        = {Server Machine Dataset ({SMD}) for {O}mnianomaly},
  howpublished = {Kaggle},
  year         = {2023},
  url          = {https://www.kaggle.com/datasets/mgusat/smd-onmiad},
  note         = {Accessed: 2026-10-08}
}

\appendixpage
\appendix

\section{Estimation of the geometric median and of the median covariation matrix} \label{appendix:geommed_mcm}

\subsection{Estimation of the geometric median}

The geometric median $m$ is an extension to $\mathbb{R}^d$ of the notion of the real median. Considering a random vector $ X $ lying in $\mathbb{R}^{d}$, the geometric median of   $X$ is defined as  \citep{haldane1948note}     
\begin{equation*}
    m = \operatorname{argmin}_{h \in \mathbb{R}^d}\mathbb{E}[\|X - h\| - \|X\|] ,
\end{equation*}
where $\| . \|$ is the Euclidean norm.
Observe that the term $-\|X\|$ in the objective function obviates the need to assume the existence of a first-order moment for $X$. In addition, to guarantee existence and uniqueness of $m$, we require that 
the random vector $X$ is not concentrated on a straight line \citep{kemperman1987median}.
Finally, it is well known that if the distribution of $X$ is symmetric around its mean $\mu$, the geometric median coincides with the mean. Unlike the mean, the geometric median does not have a known closed-form expression; however, several numerical methods are available for its estimation 
The more usual methods are an iterative one consisting  in the Weiszfeld's algorithm 
\citep{weiszfeld1937point,vardi2000multivariate} and an online method introduced by \cite{cardot2013efficient} and consisting in an averaged stochastic gradient algorithm (ASGD for short). 
These algorithms are precisely described below.  

\paragraph{Weiszfeld algorithm.}
The Weiszfeld algorithm, introduced by \cite{weiszfeld1937point} and later refined by \cite{vardi2000multivariate} and \cite{beck2015weiszfeld}, is a fixed-point iteration method for computing the geometric median. Given $X_1, \ldots , X_N$ i.i.d. $N$ copies of $X$; at iteration $t+1$, the update rule is given by:
\begin{equation}\label{eqWeizsfled}
m_{t+1} = \frac{ \sum_{k=1}^N \frac{X_k}{\|X_k - m_t\|} }{ \sum_{k=1}^N \frac{1}{\|X_k - m_t\|} }
\end{equation}
This algorithm exhibits two important properties. First, each iteration requires recomputing weights for all $N$ data points, resulting in a per-iteration complexity of $\mathcal{O}(Nd)$. Consequently, after $T$ iterations, the total computational cost scales as $\mathcal{O}(NdT)$. Second, the algorithm operates offline, meaning that incorporating new data points requires restarting the computation entirely, making it unsuitable for streaming data scenarios. 

\paragraph{Averaged stochastic gradient algorithm.}
A faster and more adaptive way, in term of computational complexity, to estimate the geometric median is given by an averaged stochastic gradient algorithm, (see \cite{robbins1951stochastic}, \cite{ruppert1988efficient},\cite{cardot2013efficient}, and \cite{godichon2016estimating}). Considering $N$ i i d copies $X_1$,...,$X_N$ arriving sequentially, it is defined recursively for all $n \geq 0$ by :  
\begin{align}\label{mediangeometric}
m_{n + 1} & = m_n + \gamma_{n + 1} \frac{X_{n + 1} - m_n}{\| X_{n + 1} - m_n \|} \\
\overline{m}_{n+1} &= \overline{m}_n - \frac{1}{n + 2}(m_{n + 1} - \overline{m}_n) \label{mediangeometricmean}
\end{align}
with $m_0 = \overline{m}_0$ chosen arbitrarily. Intuitively, the fact that the gradient norms \( \frac{X_{n + 1} - m_n}{\| X_{n + 1} - m_n \|} \) are bounded limits the influence of an outlier. The convergence of the algorithm is accelerated by the averaging operation in the second line. Indeed, the estimates may oscillate around the estimated parameter, and averaging the estimates helps accelerate the convergence. $L^p$ and almost sure rates of convergence  of $\overline{m}_n$, are provided in \cite{godichon2016estimating}, and under certain assumptions asymptotic efficiency in \cite{cardot2013efficient}. 

\subsection{Estimation of the median covariation matrix}

In the case of the mean, we have seen that we can replace it by the geometric median. In the case of the variance, there is no direct robust dispersion indicator, but we can use the Median Covariation Matrix (MCM for short) introduced by 
as well as \cite{kraus2012dispersion,cardot2017fast}. It is defined as: 
\[
V  =  \operatorname{argmin}_{M \in \mathcal{M}_d(\mathbb{R})} \mathbb{E} \left[ \left\| (X - m)(X - m)^T - M \right\|_{F} - \left\| (X - m)(X - m)^T \right\|_{F} \right]
\]  
where $ \| . \|_{F}$ is the Frobenius norm for matrices. Observe that the MCM can be seen as the geometric median of the random matrix $(X-m)(X-m)^{T}$. Then, we can do the same remarks as for the median, i.e   the term $\| (X - m)(X - m)^T \|_F$  enables  not to suppose  the existence of moment of order $2$ of $X$. In addition, the uniqueness of $V$ requires the random matrix $(X-m)(X-m)^{T}$'s distribution not to be concentrated along a one-dimensional subspace of the matrix space.
Then, as in the the case of the median, there are two methods (Weisfeld's algorithm and ASGD) for estimating iteratively or recursively the MCM. These methods are precisely described below. 

\paragraph{Weisfzeld algorithm.}
The estimation is performed after obtaining an estimate \( \widehat{m} \) of $m$ using the Weiszfeld algorithm described by \eqref{eqWeizsfled}. The Weiszfeld algorithm is then adapted as follows (see \cite{weiszfeld1937point},  \cite{vardi2000multivariate} \cite{beck2015weiszfeld} and \cite{cardot2017fast} ) :
\[
V_{t+1} = \frac{\sum_{k = 1}^N \|(X_k - \widehat{m})(X_k - \widehat{m})^T - V_t \|_F^{-1}(X_k - \widehat{m})(X_k - \widehat{m})^T} {\sum_{k = 1}^N \|(X_k - \widehat{m})(X_k - \widehat{m})^T - V_t \|_F^{-1}}
\]
where $\widehat{m}$ denotes an estimate of $m$. As with the geometric median, it is also a fixed point iteration method, needing $\mathcal{O}(N d^2T)$ computations. 

\paragraph{Averaged Stochastic Gradient Algorithm.}
As with the geometric median, it is possible to accelerate the algorithm using an averaged stochastic gradient algorithm, as defined (\cite{cardot2017fast})   :
\begin{align}
V_{n + 1} &= V_n + \gamma_{n+1} \frac{(X_{n+1} - \overline{m}_n)(X_{n+1} - \overline{m}_n)^T - V_n}{||(X_{n+1} - \overline{m}_n)(X_{n+1} - \overline{m}_n)^T - V_n||_F} \\
\overline{V}_{n+1} &= \overline{V}_n - \frac{1}{n + 2} \left( \overline{V}_n - V_{n + 1} \right) 
\end{align}
where $\overline{m}_n$ is defined by \eqref{mediangeometricmean}, and $V_0 = \overline{V}_0$ chosen arbitrarily. Under certain assumptions, convergence in distribution guarantees for \( \overline{V}_{n} \) are also provided in \cite{cardot2013efficient}.

\section{Influence functions}\label{influence_functions}

The influence function \citep[see][]{hampel1974influence} describes the impact of a small fraction $\varepsilon$ of outliers on the estimation of a parameter $T$ of a reference distribution $\mathcal{F}_0$. Denoting $\mathcal{F}_1$ the distribution of the outliers, the observations are supposed to be distributed according to the mixture distribution $(1 - \varepsilon)\mathcal{F}_0 + \varepsilon \mathcal{F}_1$, and the influence function for the parameter $T$ of a contamination according to $\mathcal{F}_1$ is defined as
$$
IF(T, \mathcal{F}_1) = 
\lim_{\varepsilon \to 0} \frac{T\big((1 - \varepsilon)\mathcal{F}_0 + \varepsilon \mathcal{F}_1\big) - T(\mathcal{F}_0)}{\varepsilon}.
$$
In particular, if $\mathcal{F}_0$ has mean $\mu_0$ and variance $\Sigma_0$, we have that $IF(\mu, \mathcal{F}_1) = \delta$ and $IF(\Sigma, \mathcal{F}_1) = \delta \delta^\top + \Sigma_1 - \Sigma_0$, where $\mu_1$ and $\Sigma_1$ stand for the mean and variance of $\mathcal{F}_1$, respectively, and $\delta = \mu_1 - \mu_0$.

The simulation setting described in Section~\ref{subsec:simdesign} combines three types of contamination: ($i$) a mean shift ($\mu_1 = \mu_0 + k m_1$), ($ii$) an inflation of the variance ($\Sigma_1 = \ell \Sigma_0$), and ($iii$) a shape transformation of the variance ($\Sigma_1 = D_0 T(\rho_1) D_0$, where $D_0$ is diagonal, $T(\rho)$ stands for the Toeplitz matrix with entries $\rho^{|i-j|}$, and $\Sigma_0 = D_0 T(\rho_0) D_0$). 

Then, the influence functions for the mean and the variance under each type of contamination are as follows:
$$
\begin{array}{lcc|cc}
     & \mu_1 & \Sigma_1 & IF(\mu, \mathcal{F}_1) & IF(\Sigma, \mathcal{F}_1) \\
     \hline
     (i) & \mu_0 + k m_1 & \Sigma_0 & k m_1 & k^2 m_1 m_1^\top \\ 
     (ii) & \mu_0 & \ell \Sigma_0 & 0 & (\ell - 1) \Sigma_0 \\ 
     (iii) & \mu_0 & D_0 T(\rho_1) D_0 & 0 & D_0 \big(T(\rho_1) - T(\rho_0)\big) D_0
\end{array}
$$
where we observe that the influence function for the mean is unbounded for $k \geq 0$, and that for the variance is unbounded only for $\ell \geq 1$. As for the shape transformation ($iii$), we see that the influence function for the variance is always bounded, whatever the value of $\rho_1$.  

\section{Deformation scenario}\label{app:deformation}

\begin{figure}[!ht]
\centering

\textbf{Deformation} \\[1em]

\begin{tabular}{c c}
&
\begin{tabular*}{0.95\textwidth}{@{\extracolsep{\fill}}ccccc@{}}
\textbf{\hspace{0.3cm}\shortstack{Frobenius \\ norm error}} &
\textbf{FNR} &
\textbf{FPR} &
\textbf{AUC\hspace{0.5cm}} &
\textbf{ACC \hspace{0.5cm}}
\end{tabular*}
\\[1em]
\raisebox{1.25\height}{\rotatebox{90}{$\rho_1 = 0.5$}}
&
\includegraphics[width=0.95\textwidth,height=4cm]{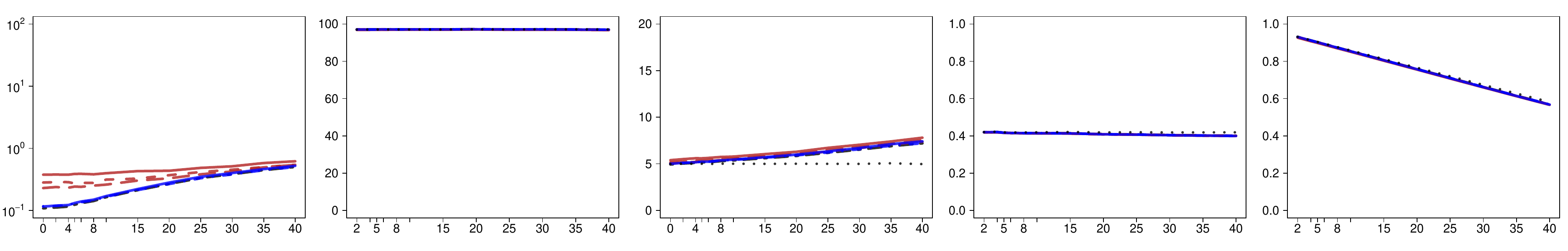}
\\[1em]

\raisebox{1.25\height}{\rotatebox{90}{$\rho_1 = 0.7$}}
&
\includegraphics[width=0.95\textwidth,height=4cm]{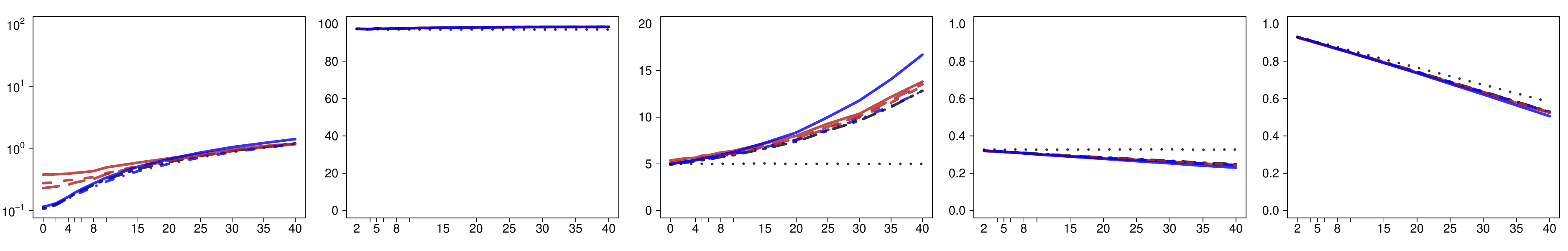}
\\[1em]

\raisebox{1.25\height}{\rotatebox{90}{$\rho_1 = 0.95$}}
&
\includegraphics[width=0.95\textwidth,height=4cm]{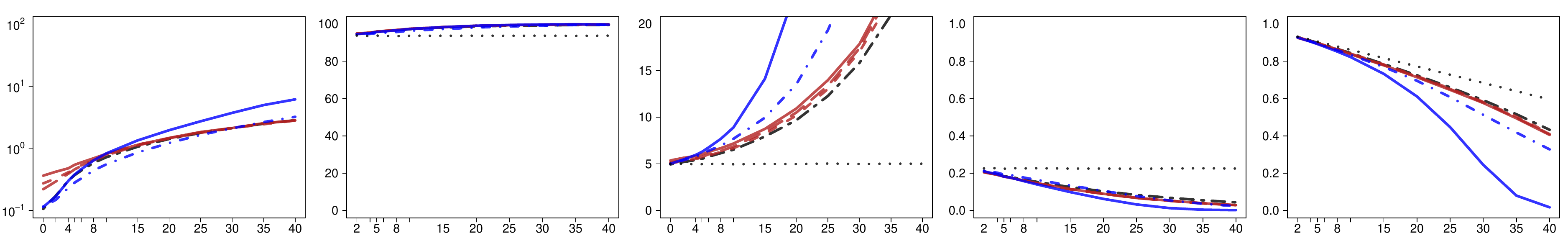}
\\
\end{tabular}

\caption{
  Evaluation criteria, Frobenius norm error, FNR, FPR, ACC and AUC, as a function of the contamination rate for the \textbf{deformation} scenario, with increasing values of $\rho_1$. \textit{Legend: same as Figure~\ref{fig:dilatation}.
}
}
\label{fig:deformation}
\end{figure}

The deformation scenario modifies the orientation of the covariance structure 
through a correlation parameter $\rho_1$, while preserving its overall scale. As illustrated in Figure~\ref{fig:deformation}, this regime shares the same qualitative behavior as concentration scenario: outliers remain close to the bulk, ACC 
and AUC stay low, and the ranking of the methods is unchanged. \textit{FAST-MCD} and \textit{OGK} both exhibit the same bias as under concentration, for identical reasons. \textit{FAST-MCD} assumes a bounded contamination fraction and distant outliers, while \textit{OGK}'s pairwise estimators are affected by correlations between the outliers and the bulk. The proposed online and streaming methods again adapt their threshold to the empirical distribution of Mahalanobis distances and produce fewer false 
discoveries than \textit{OGK} and \textit{MCD}, despite an FPR above the nominal 5\% level.

\section{Data preprocessing}\label{app:preprocessing}

Prior to the analysis, the SMD data are preprocessed consistently across all 
machines to improve numerical stability and ensure comparability across methods. 
The preprocessing pipeline consists of four steps.

\paragraph{Variable filtering.}
Variables containing non-finite values or too few unique values are removed, as 
they provide limited information for anomaly detection and may lead to unstable 
covariance estimation. Variables with near-zero empirical standard deviation or 
MAD are also discarded to avoid degenerate or ill-conditioned covariance matrices.

\paragraph{Redundancy removal.}
To mitigate feature redundancy and near-linear dependencies, we filter highly correlated variables using a correlation threshold of $0.7$.

\paragraph{Robust standardization.}
The remaining variables are standardized using their median and MAD, so that all 
variables are placed on a comparable scale, preventing variables with larger 
magnitudes or dispersion from dominating covariance- and Mahalanobis-based 
anomaly scores. The use of median and MAD also reduces the influence of 
anomalous observations during scaling.

\paragraph{Batch size.}
The streaming procedures use a fixed batch size of $10$ for every machine.

\end{document}